\pdfoutput=1
\documentclass[11pt,twoside,a4paper,cmspaper,final,collab]{cms-tdr}
\def\svnVersion{b0ff8dc-D}\def\svnDate{2021/02/20}\def\cmsCernNoTag{CERN-EP-2020-188}\def\cmsCernDate{\today}\def\cmsMessage{"Published in the Journal of High Energy Physics as \href{http://dx.doi.org/10.1007/JHEP04(2021)124}{\doi{10.1007/JHEP04(2021)124}.}"}
\begin{document}\cmsNoteHeader{BPH-15-009}

\newlength\cmsTabSkip\setlength{\cmsTabSkip}{1ex}

\newcommand{\qq}{\ensuremath{q^2}\xspace}
\newcommand{\lppi}{\ensuremath{\PgL \to \Pp \Pgpm}\xspace}
\newcommand{\Kst}{\ensuremath{\PK^{\ast+}}\xspace}
\newcommand{\Kstmass}{\ensuremath{\PK^{\ast}(892)^+}\xspace}
\newcommand{\afb}{\ensuremath{A_\mathrm{FB}}\xspace}
\newcommand{\fl}{\ensuremath{F_\mathrm{L}}\xspace}
\newcommand{\as}{\ensuremath{A_\mathit{S}}\xspace}
\newcommand{\fs}{\ensuremath{F_\mathit{S}}\xspace}
\newcommand{\Sig}{\ensuremath{\PBp\to \Kst \Pgmp \Pgmm}\xspace}
\newcommand{\JPS}{\ensuremath{\PBp\to \Kst \JPsi }\xspace}
\newcommand{\PSP}{\ensuremath{\PBp\to \Kst \Pgy }\xspace}

\cmsNoteHeader{BPH-15-009} 
\title{Angular analysis of the decay \texorpdfstring{$\PBp\to \Kstmass \Pgmp \Pgmm$}{B+ to K*+ mu+ mu-} in proton-proton
collisions at \texorpdfstring{$\sqrt{s} = 8\TeV$}{sqrt(s) = 8 TeV}}

\date{\today}

\abstract{Angular distributions of the decay $\PBp\to \Kstmass \Pgmp \Pgmm$ are studied using 
events collected with the CMS detector in $\sqrt{s}=8\TeV$ 
proton-proton collisions at the LHC, corresponding to an integrated luminosity of 20.0\fbinv.
The forward-backward asymmetry of the muons and the longitudinal polarization
of the \Kstmass meson are determined as a function of the square of the dimuon
invariant mass. These are the first results from this exclusive decay mode and are in agreement with
a standard model prediction.}

\hypersetup{
pdfauthor={CMS Collaboration},
pdftitle={Angular analysis of the decay B+ to K*(892)+ mu+ mu- in proton-proton collisions at sqrt(s) = 8 TeV},
pdfsubject={CMS},
pdfkeywords={CMS, B physics}}

\maketitle 

\section{Introduction}
The decays of heavy-flavor hadrons can be used to probe high mass scales by searching for effects caused by unknown
heavy particles that modify the standard model (SM) description of the decay.  Flavor changing neutral current
decays, such as those involving $\PQb\to \PQs \mu^+ \mu^-$ transitions, are particularly promising as they are forbidden at tree
level, and only occur via loop diagrams.  The lack of a dominating tree-level process allows for a greater sensitivity
to the effects of new particles. These effects can appear as differences
in the overall decay rate or as modifications to the angular distributions of the decay products.

In this paper, an analysis of the \Sig decay is performed, where \Kst indicates the \Kstmass meson.  Charge-conjugate
states are implied throughout the paper.  The theoretical description of this decay requires four independent kinematic
variables, which are chosen by convention to be three angles plus the square of the dimuon invariant mass (\qq).  Two
angular distributions are used to measure two decay observables, the muon forward-backward asymmetry, \afb, and the \Kst
longitudinal polarization fraction, \fl, in bins of \qq.  The data for this analysis were collected in proton-proton
$(\Pp\Pp)$ collisions at a center-of-mass energy of $8\TeV$ by the CMS detector at the CERN LHC, and correspond to an integrated
luminosity of 20.0\fbinv~\cite{LUMI}.  Previous measurements of \afb and \fl have been made in the exclusive mode
$\PBz\to\PKstP{892}{}^0 \Pgmp \Pgmm$~\cite{CDF, LHCb, Chatrchyan:2013cda, Kst0mumu, Aaij:2015oid, Aaij:2020nrf, Aaboud:2018krd}
and in a combination of decays
of the form $\PB\to\PKstP{892}\ell^+\ell^-$~\cite{BaBar, Belle, Lees:2015ymt}, where $\ell$ refers to an electron or a muon and the
combinations are of $\PKstP{892}$ isospin states and/or lepton flavor states.
The results are generally consistent with the SM predictions~\cite{Bobeth:2010wg, Bobeth:2011nj,
Bobeth:2012vn, Ali:2006ew, Altmannshofer:2008dz, Altmannshofer:2011gn, Jager:2012uw, Descotes-Genon:2013vna,
Descotes-Genon:2014uoa, Alguero:2018nvb, Alguero:2019ptt}.
This paper reports the first measurement of \afb and \fl in the exclusive decay \Sig, with the
\Kst meson reconstructed in the $\PKzS\PGpp$ decay mode and the \PKzS meson identified from its decay to a pair of charged pions.

\section{The CMS detector}
The central feature of the CMS apparatus is a superconducting solenoid of 6\unit{m} internal diameter, providing a
magnetic field of 3.8\unit{T}. Within the solenoid volume are a silicon pixel and strip tracker, a lead tungstate
crystal electromagnetic calorimeter, and a brass and scintillator hadron calorimeter, each composed of a barrel and two
endcap sections. The silicon tracker measures charged particles within the pseudorapidity range $\abs{\eta} <
2.5$. During the LHC running period when the data used in this paper were recorded, the silicon tracker consisted of
1440 silicon pixel and 15\,148 silicon strip detector modules. For non-isolated particles of $1 < \pt < 10\GeV$ and
$\abs{\eta} < 1.4$, the track resolutions are typically 1.5\% in \pt and 25--90 (45--150)\mum in the transverse
(longitudinal) impact parameter~\cite{TRK-11-001}.  Muons with $\abs{\eta} < 2.4$ are measured with gas-ionization
chambers embedded in the steel flux-return yoke outside the solenoid. A more detailed description of the CMS detector, together with a definition of
the coordinate system used and the relevant kinematic variables, can be found in Ref.~\cite{CMS}.  Distances that are measured
with respect to the beamline are in the transverse plane.

Events of interest are selected using a two-tiered trigger system~\cite{Khachatryan:2016bia}. The first level, composed of custom
hardware processors, uses information from the calorimeters and muon detectors.
The second level, known as the high-level trigger, consists of a farm of processors running a version of the full event
reconstruction software optimized for fast processing.

\section{Event selection}
The events used in this analysis are selected by a trigger designed specifically for finding \PQb hadron decays
that include two muons.  The trigger requires two oppositely charged muons, each with transverse momentum $\pt > 3.5\GeV$ and
$\abs{\eta}<2.2$.  The two muons are fitted to a common vertex and retained if the fit $\chi^2$ probability is greater than
10\% and the vertex is displaced from the beamline by at least three times the 
uncertainty in the distance.  
The dimuon system is further required to have $\pt > 6.9\GeV$, invariant mass between 1 and 4.8\GeV, and a momentum 
vector whose angle $\alpha$ with respect to the vector between the beamline and the dimuon vertex 
satisfies $\cos\alpha > 0.9$.

The offline reconstruction of the signal decay \Sig requires two oppositely charged muons
and a \Kst meson, where the \Kst meson is reconstructed in the $\PKzS\PGpp$ decay mode, and the $\PKzS$ meson is identified through its decay to
$\PGpp\PGpm$.  The trigger requirements are reapplied to the corresponding offline quantities and the offline
muon candidates must pass the soft muon criteria~\cite{Chatrchyan:2012xi} and correspond to the muons that satisfied
the trigger requirements.  The \PKzS meson candidates are
reconstructed by fitting pairs of oppositely charged tracks to a common vertex and selected using standard selection criteria.
In particular, the tracks must have at least 6 hits
in the silicon tracker, a $\chi^2$ per degree of freedom (dof) less than 5, pass at a distance from the beamline at
least 2 times its uncertainty, and have the closest distance between their trajectories be less than 1\cm.  In addition, the
fitted vertex must have a $\chi^2/\text{dof}<7$ and be located at a distance from the beamline that is at least 15 times the
calculated uncertainty in the distance. The
two-track invariant mass must be within 17.3\MeV (three times the average resolution) of the \PKzS meson mass~\cite{PDG} when the
tracks are assigned the charged pion mass.  To remove \lppi decays, the two-track combination is rejected if the
invariant mass is in the range 1.11--1.125\GeV when the high and low momentum tracks are assigned the proton and charged pion
mass, respectively.  Each \PKzS candidate is combined with two oppositely charged muons and a non-muon track, assumed to be a pion,
in a fit to a common vertex to form a \PBp meson candidate.  The $\PKzS\PGpp$ invariant mass is required to be within
100\MeV of the world-average \Kst mass~\cite{PDG}, and the invariant mass of the $\PKzS\PGpp\PGmp\PGmm$ system, $m$, must 
be in the range $4.76<m<5.8\GeV$.

The remaining selection criteria are obtained by maximizing $S/\sqrt{S+B}$ for different event shape variables. The
number of signal events, $S$, is obtained from the simulation (normalized to the data) and the number of background 
events, $B$, is obtained from the $\PKzS\PGpp\PGmp\PGmm$ data sideband invariant mass regions 4.76--5.18 and 5.38--5.8\GeV.  
The \PKzS meson \pt must be greater than $1\GeV$.  The pion track from the \Kst decay must have $\pt>0.4\GeV$ and an
impact parameter with respect to the beamline of at least 0.4 times the uncertainty in this parameter found from the vertex fit.
The \PBp candidate vertex must have a fit $\chi^2$ probability larger than 10\% and a
separation from the beamline of at least $12$ times the calculated uncertainty in the separation.  The angle $\alpha$
between the vector from the beamline to the vertex location and
the \PBp candidate momentum vector (in the transverse plane) must satisfy $\cos\alpha>0.9994$.  In 0.3\% of the events in which
a candidate passes the selection criteria, a second candidate also passes the same criteria.  In these cases, the candidate
with the smaller vertex fit $\chi^{2}$ value is chosen.

The decay modes \JPS and \PSP, followed by the dimuon decays of charmonium states $\JPsi$ and $\Pgy$, have the same final-state particles 
as the signal mode. As described in Section~\ref{sec:Analysis}, the analysis is performed in bins of \qq that exclude candidates in 
the \JPS and \PSP regions, namely $8.68 < \qq < 10.09 \GeV^2$ and $12.86 < \qq < 14.18 \GeV^2$.
However, since events from charmonium decay are produced quite copiously, a significant contribution can still appear
in the signal \qq regions.  This primarily occurs through two effects:
finite detector resolution resulting in a reconstructed dimuon mass different than the true value, and decays of the
two charmonium states in which a low-energy photon is emitted in addition to the two muons.  Two additional requirements are used
to remove these contributions.  First, candidates that 
satisfy either $m_\JPsi - 5\sigma_q < q < m_\JPsi + 3\sigma_q$ or $\abs{q-m_\Pgy} < 3\sigma_q$ are removed,
where $m_\JPsi$ and $m_{\Pgy}$ are the world-average \JPsi and \Pgy masses~\cite{PDG}, respectively, and $\sigma_{q}$ is 
the calculated uncertainty in $q$ for each candidate.  The second requirement specifically targets the radiative
background by using the fact that the missing low-energy photon will shift $q$ and $m$ from their nominal values by a similar
amount.  Thus, these events are suppressed by requiring
$\smash[b]{\abs{(m-m_\PBp)-(q-m_\JPsi)} > 0.09\GeV}$ and $\smash[b]{\abs{(m-m_\PBp)-(q-m_\Pgy)} > 0.03\GeV}$\@.
When the \JPS decay mode is used as a control sample, the requirements in this paragraph are not applied.
\begin{figure}
\centering
    \includegraphics[width=0.9\textwidth]{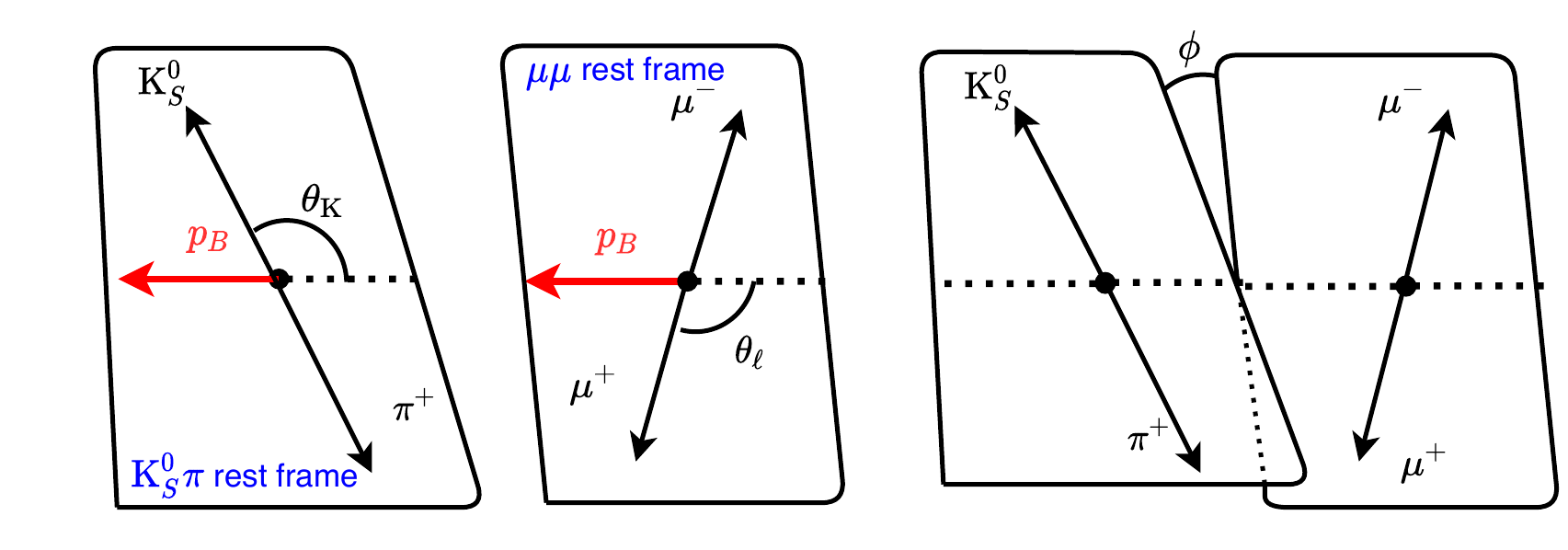}
      \caption{Definition of the angular observables $\theta_{\PK}$ (left), $\theta_{\ell}$ (middle), and $\phi$ (right) for the decay \Sig.}
    \label{fig:axis}
\end{figure}

The Monte Carlo (MC) samples corresponding to the signal and control channels are simulated
using \PYTHIA6.426~\cite{Pythia}, with the unstable particle decays modeled by \EVTGEN~\cite{EvtGen}.
The particles are then propagated through a detailed model of the CMS detector with \GEANTfour~\cite{Geant4}.
The reconstruction and selection of the MC generated events follow the same algorithms as for the collision data. 
The number and spatial distribution of additional $\Pp\Pp$ collision vertices in the same or nearby beam crossings
in the data are simulated by weighting the MC samples to match the distributions found in data.
The signal MC samples are used to estimate
the efficiency, which includes the detector acceptance, the trigger efficiency, and the efficiency for reconstructing and selecting the
signal candidates.

\section{Angular analysis}
\label{sec:Analysis}

The measurement of \afb and \fl is performed in three \qq regions: $1 < \qq < 8.68\GeV^{2}$, $10.09 < \qq < 12.86\GeV^{2}$,
and $14.18 < \qq < 19\GeV^{2}$. The angular distribution of the signal process, \Sig, depends on three 
variables as shown in Fig.~\ref{fig:axis}: $\theta_\PK$ (the angle in the \Kst meson rest frame between the momentum of the \PKzS meson 
and the negative of the \PBp meson momentum), $\theta_\ell$ (the angle in the dimuon rest frame between the momentum of the positively 
charged muon and the negative of the \PBp meson momentum), and $\phi$ (the angle in the \PBp meson rest frame between the plane containing
the two muons and the plane containing the $\PKzS$ and $\Pgpp$ mesons). Since the extracted angular observables \afb and \fl do not
depend on $\phi$, this angle is integrated out.  While the $\PKzS\Pgpp$ invariant mass is required to be consistent
with coming from a \Kst resonance decay, there can still be $S$-wave $\PKzS\Pgpp$
contributions~\cite{Descotes-Genon:2013vna,Becirevic:2012dp,Matias:2012qz,Blake:Swave}. This is parameterized by two
terms: the $S$-wave fraction, \fs, and the interference amplitude, \as, between $S$- and $P$-wave decays.  The parameters \afb, \fl, \fs, and \as are functions of \qq.  The differential decay rate of the signal decay \Sig, as a function of the angular variables and \qq,
can be written~\cite{Descotes-Genon:2013vna,Blake:Swave} as:

\begin{linenomath*}
\begin{equation}\label{eqn:angle}
  \begin{split}
     \frac{1}{\Gamma}\frac{\rd{}^3\Gamma}{\rd{}\!\cos\theta_\PK\, \rd{}\!\cos\theta_\ell\, \rd{}q^2}  &=
      \frac{9}{16} \left\lbrace \frac{2}{3} \Bigl[ \fs + 2\as \cos\theta_\PK \Bigr] \left(1 - \cos^2\theta_\ell\right) \right. \\
    & \left. +\; \left(1 - \fs\right) \Bigl[2 \fl \cos^2\theta_\PK \left(1 - \cos^2\theta_\ell\right) \right. \\
    & \left. +\; \frac{1}{2} \left(1 - \fl\right) \left(1 - \cos^2\theta_\PK\right) \left(1 + \cos^2\theta_\ell\right) \right. \\
    & \left. +\; \frac{4}{3} \afb \left(1 - \cos^2\theta_\PK\right) \cos\theta_\ell\Bigr] \right\rbrace.
   \end{split}
\end{equation}
\end{linenomath*}
For each \qq bin, the observables \afb and \fl are extracted by performing an unbinned extended maximum likelihood fit
with three independent variables: $m$, $\cos\theta_\PK$, and $\cos\theta_\ell$. The unnormalized probability density function (pdf) used to
fit the data is:
\begin{linenomath*}
\begin{equation} \label{eqn:pdf}
  \begin{split}
    \mathrm{pdf}(m,\cos\theta_\PK,\cos\theta_\ell) & = Y_S\, S^m(m) \, S^a(\cos\theta_\PK,\cos\theta_\ell)\, \epsilon(\cos\theta_\PK,\cos\theta_\ell)  \\
    & + Y_B\, B^m(m)\, B^{\theta_\PK}(\cos\theta_\PK)\, B^{\theta_\ell}(\cos\theta_\ell).
  \end{split}
\end{equation}
\end{linenomath*}
The parameters $Y_S$ and $Y_B$ are the signal and background yields, respectively, and are free parameters in the fit.
The signal mass shape, $S^m(m)$, is modeled by the sum of two Gaussian functions with a common mean,
and the shape parameters are fixed to the values obtained from fitting simulated signal events.  
The mass shape of the background, $B^m(m)$, is an exponential function
with the exponent as a free parameter.  The function $S^a(\cos\theta_\PK,\cos\theta_\ell)$ is obtained from
Eq.~(\ref{eqn:angle}) to describe the signal event distribution in the $(\cos\theta_\PK,\cos\theta_\ell)$ angular space.
Since the $S$-wave contribution is found to be small, \fs and \as are fixed to zero in the nominal fit. The functions
$B^{\theta_\PK}(\cos\theta_\PK)$ and $B^{\theta_\ell}(\cos\theta_\ell)$ are the background shapes in the angular
space. They are obtained by fitting the data events in the \PBp invariant mass sideband regions and fixed in the final fit.
The $B^{\theta_\PK}(\cos\theta_\PK)$ distributions are fitted to a sum of two exponential functions, a fourth-degree polynomial, and a third-degree 
polynomial for the low, middle, and high \qq ranges, respectively. 
The $B^{\theta_\ell}(\cos\theta_\ell)$ distributions are fitted to a sum of two Gaussian functions, a fourth-degree polynomial, and a 
linear function for the low, middle, and high \qq ranges, respectively.

\begin{figure}[hbtp]
\centering
    \includegraphics[width=0.32\textwidth]{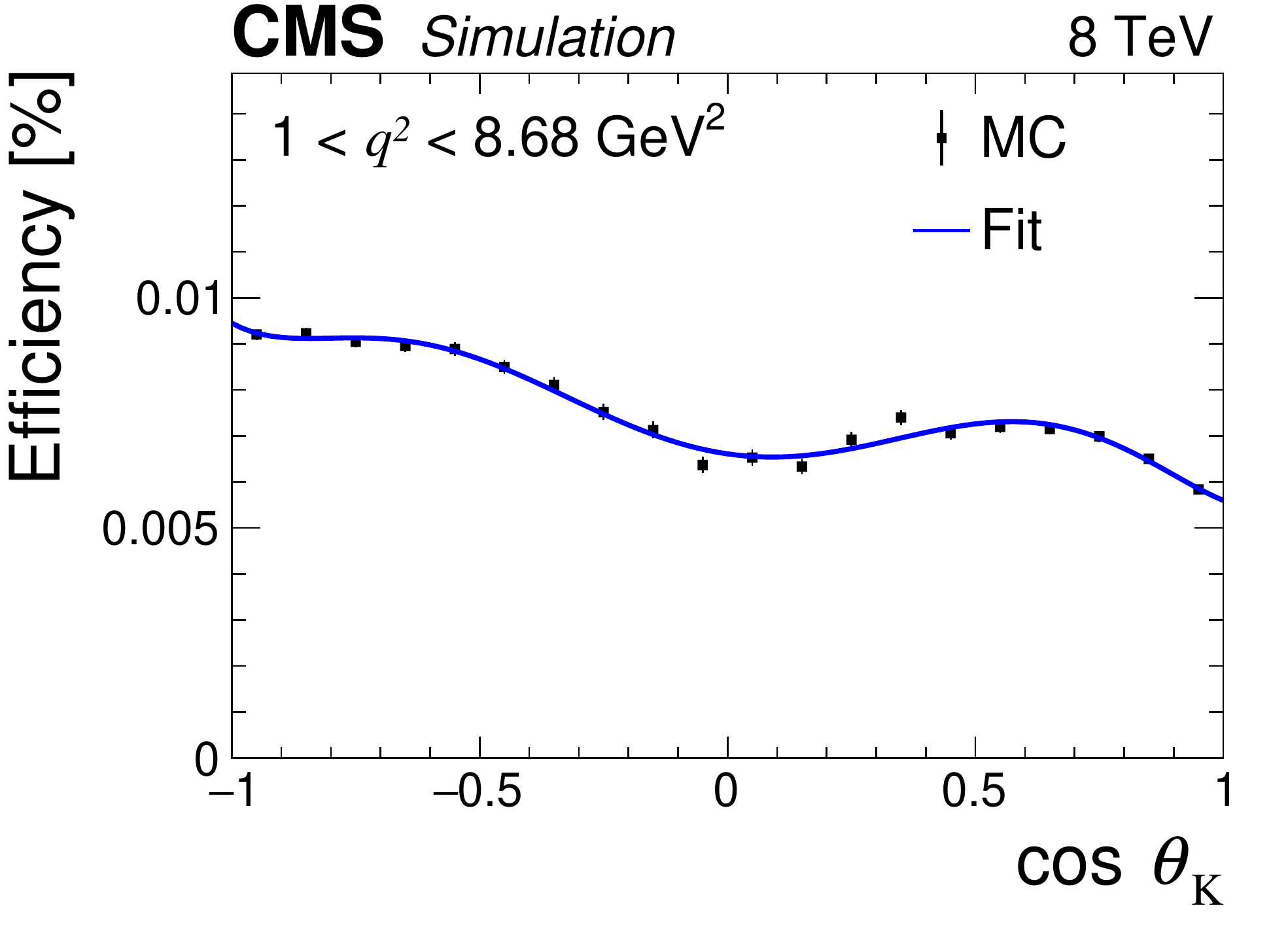}
    \includegraphics[width=0.32\textwidth]{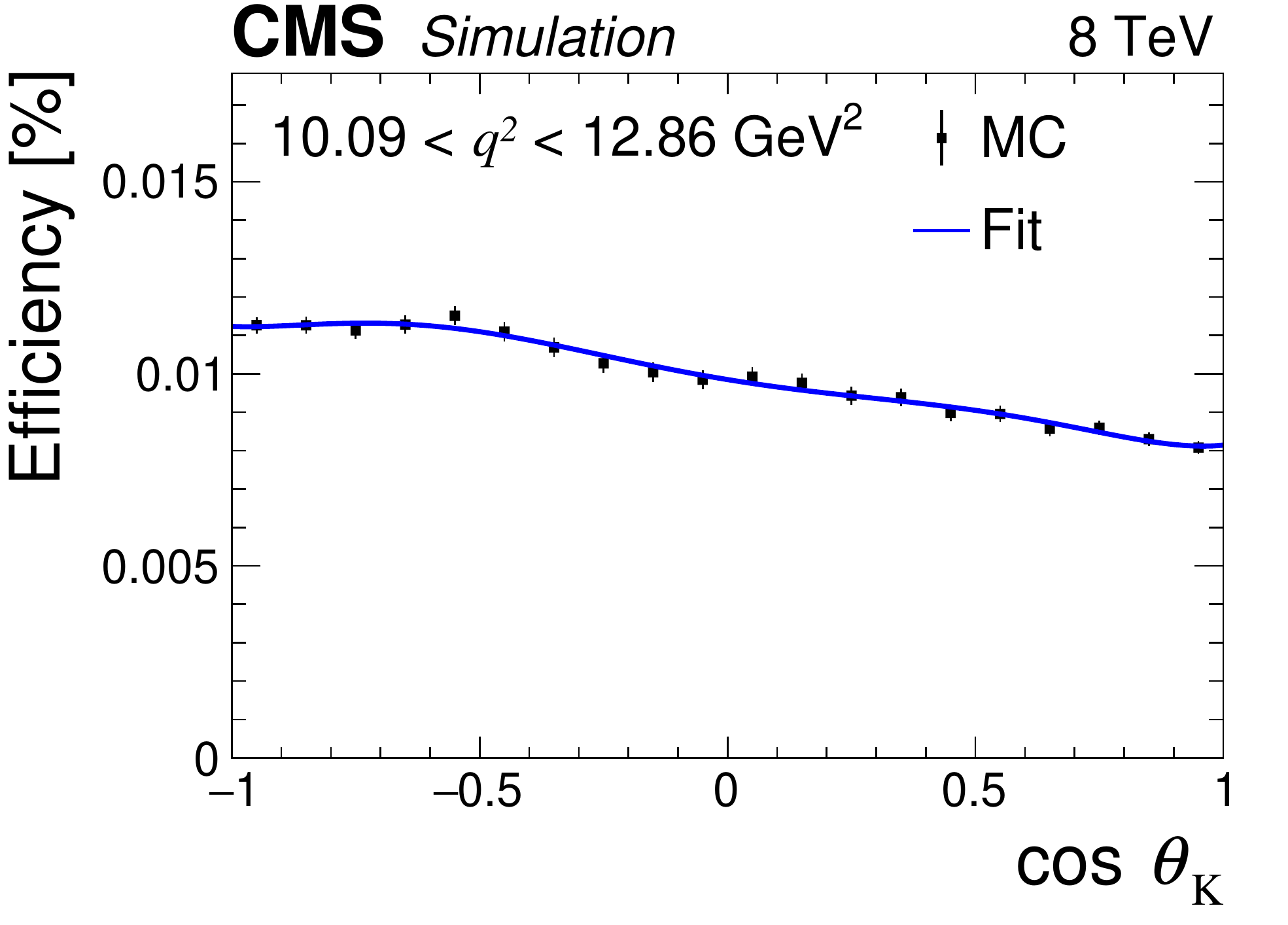}
    \includegraphics[width=0.32\textwidth]{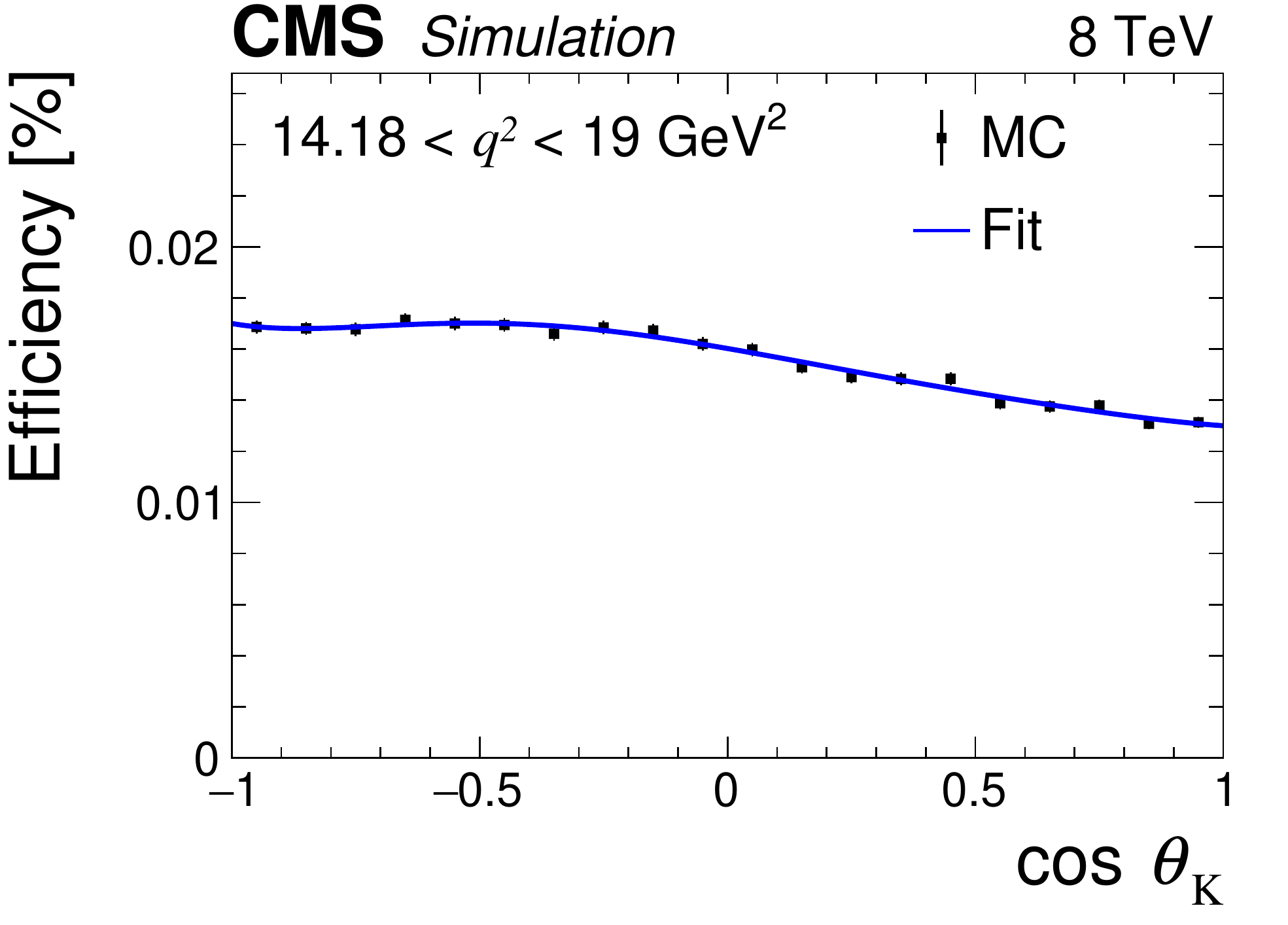}
    \includegraphics[width=0.32\textwidth]{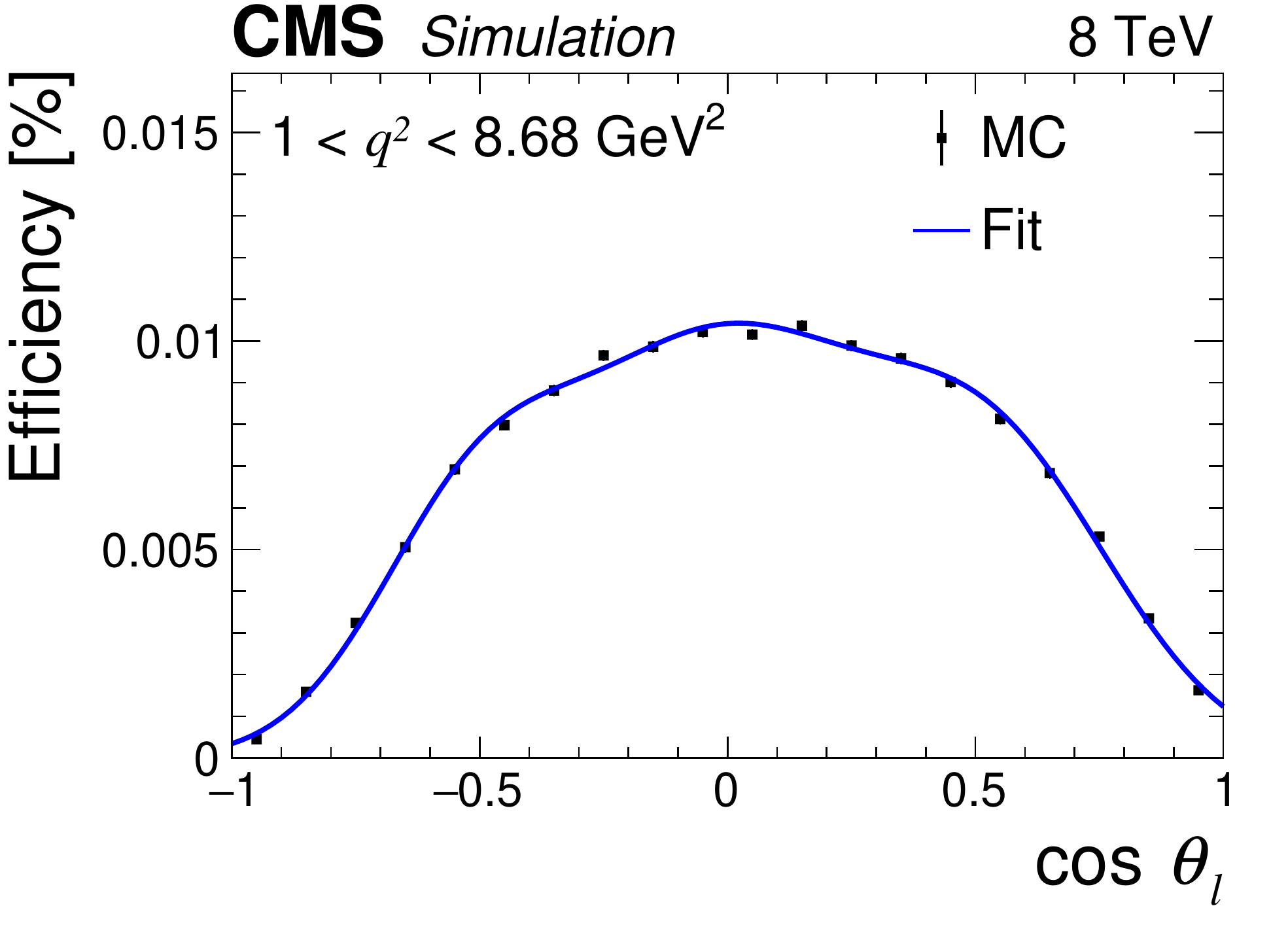}
    \includegraphics[width=0.32\textwidth]{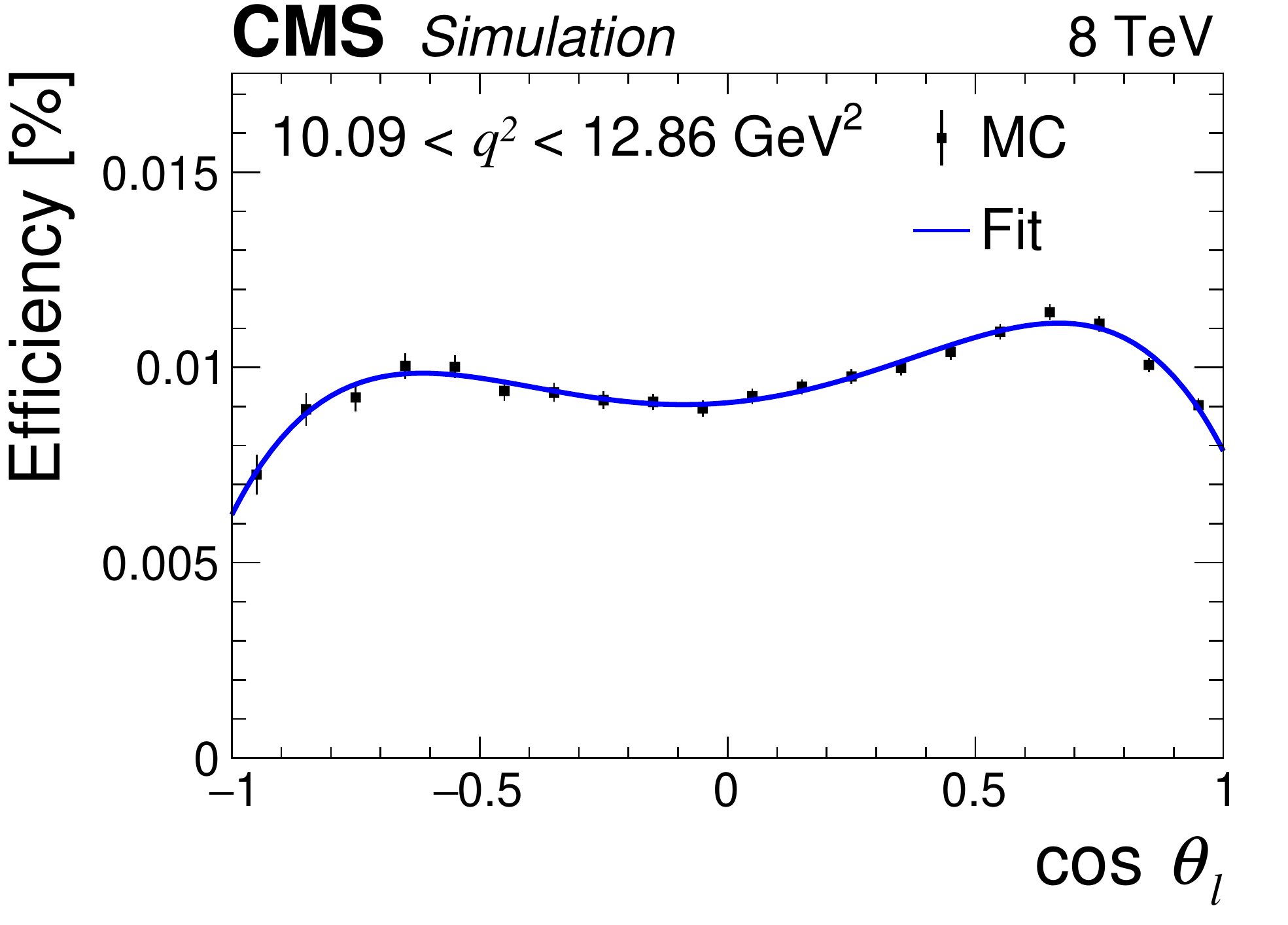}
    \includegraphics[width=0.32\textwidth]{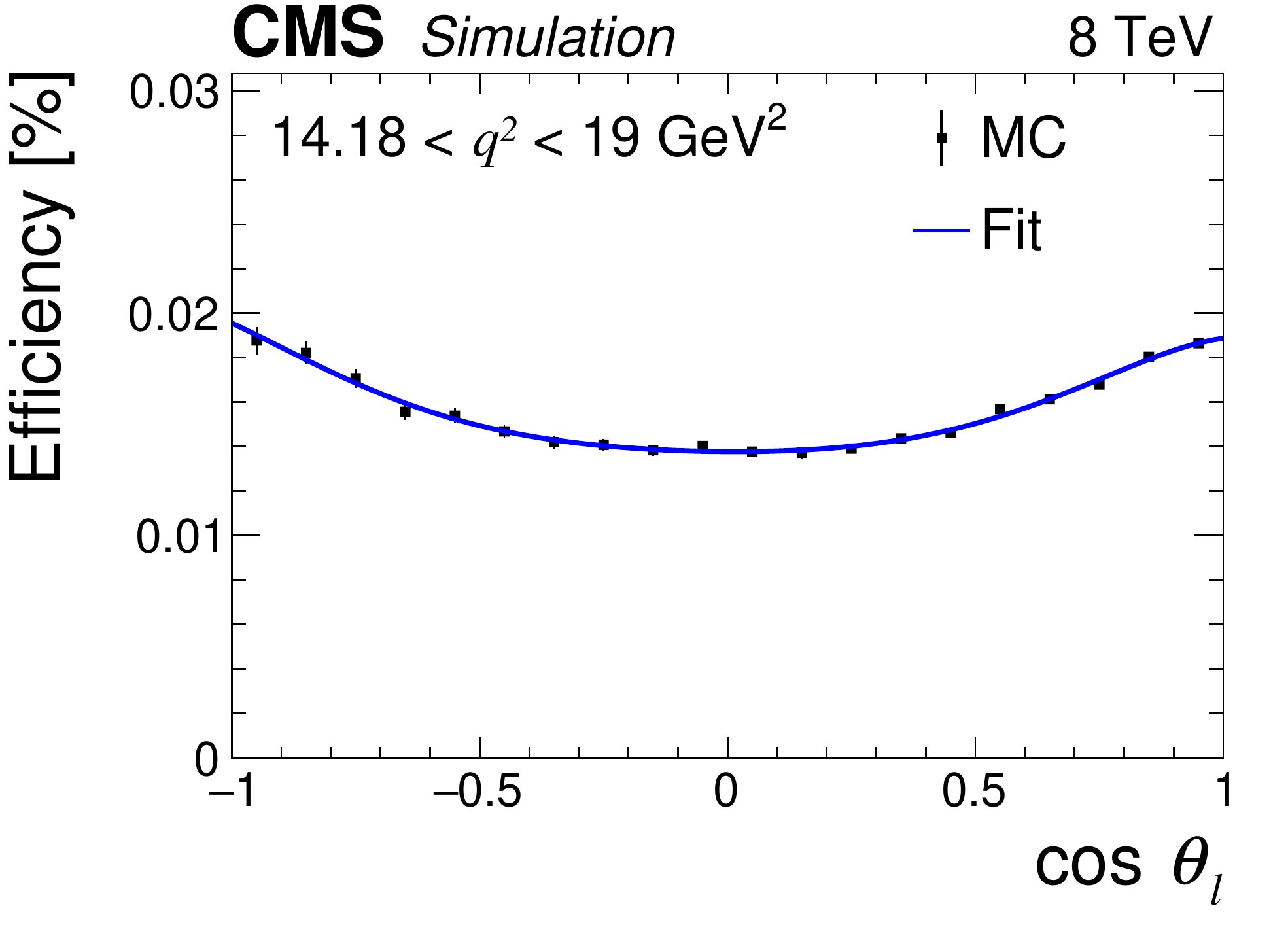}
    \caption{The signal efficiency as a function of $\cos\theta_\PK$ (upper row) and $\cos\theta_\ell$ (lower row) from simulation for the \qq ranges indicated.
    The vertical bars indicate the statistical uncertainty. The curves show the projection of the fitted result obtained from the two-dimensional fit, as described in the text.}
    \label{fig:effProjCosTheta}
\end{figure}

The signal efficiency function in the two-dimensional angular spaces $\epsilon(\cos\theta_\PK,\cos\theta_\ell)$ is
obtained from the simulated samples using a two-step unbinned maximum likelihood fit process.
In the first step, the efficiency in each \qq bin
is fitted to a product of two one-dimensional functions, one for each angular variable, assuming there is no
correlation between the variables.  The one-dimensional functions are polynomials of degree six, except for the
$\cos\theta_\ell$ distribution of the first \qq bin, which is a sum of three Gaussian functions. In the second step, 
a two-dimensional fit is performed on both angular variables, where the results from the first step are fixed, 
and an additional function is added to account for correlations.
This function is the product of the powers 0, 1, 2, and 3 for Legendre polynomials with $\cos\theta_\PK$ as the argument
and the powers 0, 1, 3, and 4 for ordinary polynomials with $\cos\theta_\ell$ as the argument.
This results in sixteen terms, each controlled by a free parameter in the fit. The signal efficiencies 
and the corresponding fits for each \qq bin are shown as projections on
$\cos\theta_\PK$ (upper plots) and $\cos\theta_\ell$ (lower plots) in Fig.~\ref{fig:effProjCosTheta}.

To test the fit, the reconstructed signal MC data set is split into 2000 random, disjoint samples, each with a similar
number of signal events as the data sample.  These are combined with background events generated using the appropriate
pdf in Eq.~(\ref{eqn:pdf}), with parameters taken from the fit to the data.  Each sample is fitted in the same manner as
the data and the resulting values for \afb and \fl are found to have approximately Gaussian distributions with mean
values close to the MC values.  This indicates the fit is unbiased and accurate, even in the presence of background.

The degree to which the simulation describes the data is examined by using the \JPS MC sample to determine the efficiency,
correcting the \JPS data by this efficiency, and comparing the $\cos\theta_\PK$ and $\cos\theta_\ell$ distributions with the SM expectations.
The residual discrepancies are found to have a negligible effect on the measured values of \afb and \fl.

\section{Systematic uncertainties}
Several sources of systematic uncertainties are considered in this analysis.
First, the statistical uncertainty associated with the finite number of signal MC events is evaluated by generating 200
alternative efficiency functions, varying the function parameters according to their uncertainties.  Each of these
efficiency functions is used to fit the data, and the standard deviations of the distributions of the fitted values for \afb and \fl
are taken as the systematic uncertainty in each quantity.  The second source of
systematic uncertainty is from the shape used to parameterize the efficiency.  The difference between the
values of \afb and \fl obtained from fitting the generator-level MC signal events (with no efficiency function) and the
reconstructed MC signal events (with the efficiency function) is taken as the estimate for this systematic uncertainty.

The third systematic uncertainty arises from modeling the angular distribution of the background events and is composed of three components. 
The first component is intended to check the functional form.  Instead of fitting the sideband data with the functional forms
described in Section~\ref{sec:Analysis}, the lower and upper sidebands are individually fit to a non-parametric function and
the two pdfs are combined according to their relative yields.
The difference between the results obtained with this alternative background pdf and the default function is taken
as a systematic uncertainty. The second component is intended to account for the uncertainty regarding how well the background in the sideband
regions represents the background in the signal region.  In the nominal fit, large \PBp invariant mass sideband regions are used to determine
the background shape in order to reduce the statistical uncertainty.  As an alternate method, the background shape is determined
from narrower sideband regions ($4.96 < m < 5.18\GeV$ and $5.38 < m < 5.6\GeV$), which are expected to be more representative of the signal
region.  Once the new background shape is determined, the fit is redone using all events (including the original sideband
region), and the change in \afb and \fl with respect to the nominal fit is used as the systematic uncertainty.  Since the
background shape parameters are fixed in the determination of \afb and \fl, the third component
accounts for the statistical uncertainty in the background shape.  The data are fitted with 200 different background shapes obtained by
varying the shape parameters by their uncertainties. The standard deviation of the distributions of the angular
observables \afb and \fl obtained from these 200 fits is included as a systematic uncertainty. 

The fourth source of systematic uncertainty is the effect from $S$-wave contamination.
The nominal fit does not include any $S$-wave contribution.  We perform an alternative fit in which the
$S$-wave fraction \fs is set to 5\% and the $S$-$P$ interference term \as is a free parameter.  The change in \afb and \fl
from the default fit is taken as the systematic uncertainty from $S$-wave contamination.  Since the analysis of the
similar decay mode $\PBz \to \PK^{\ast 0} \Pgmp \Pgmm$ did not find \fs above 3\% in any \qq bin with many more signal
events~\cite{Kst0mumu}, an upper limit of 5\% is a conservative choice.

The total systematic uncertainty is obtained by adding the individual contributions in quadrature for each \qq bin.  
The systematic uncertainties, all considered to be symmetric, are summarized in Table~\ref{tab:sys}.

\begin{table}[htbp]
\centering
    \topcaption{Sources of systematic uncertainties and the effect on \afb and \fl. The values given are absolute and the ranges
      indicate the variation over the \qq bins.\label{tab:sys}}
    \begin{tabular}{lr@{~--~}lr@{~--~}l}
      Source & \multicolumn{2}{c}{\afb $(10^{-3})$} & \multicolumn{2}{c}{\fl $(10^{-3})$} \\[1pt]
      \hline
      MC statistical uncertainty & 12 & 29 & 18 & 38 \\
      Efficiency model & 3 & 25 & 4 & 12 \\
      Background shape functional form & 0 & 9 & 0 & 33 \\
      Background shape statistical uncertainty & 16 & 73 & 20 & 87 \\
      Background shape sideband region & 28 & 153 & 38 & 78 \\
      $S$-wave contamination & 4 & 22 & 5 & 12 \\[\cmsTabSkip]
      Total systematic uncertainty & 42 & 174 & 55 & 127 \\
    \end{tabular}
\end{table}

\section{Results}
Fits to the data are performed in three independent \qq bins between 1 and 19$\GeV^2$.
As described in Section~\ref{sec:Analysis}, the measured values for \afb and \fl are obtained from an unbinned maximum likelihood
fit in which both parameters are allowed to vary freely.  The necessity of a nonnegative decay rate results in physical limits
on \afb and \fl that make it difficult to determine the statistical uncertainties from the likelihood function. Therefore, the one
dimensional uncertainty for \afb, and separately for \fl, are evaluated using Neyman constructions 
following the method of Feldman-Cousins~\cite{FC97}, generalized to treat nuisance parameters in the test statistic by the profile
likelihood method. In the construction for \afb, \fl is included in the nuisance parameters, and vice versa.  
In the Monte Carlo simulation of pseudo-experiments for obtaining the acceptance intervals in the construction, 
the nuisance parameters are treated by a parametric bootstrap procedure with profiling.  
That is, for each test value of the parameter of interest, the model including nuisance parameters is fit to the data 
to obtain the values of nuisance parameters that are used in the pseudo-experiments for constructing the acceptance 
intervals for that test value of the parameter of interest. The correlation coefficients between the two angular observables 
returned by \textsc{minuit}~\cite{Minuit} are found to be 0.1 or less, depending on the \qq bin. Tests with pseudo-experiments 
are used to verify that the statistical uncertainties have a coverage exceeding 68.3\% in all cases.

\begin{figure}[htbp]
    \centering
    \includegraphics[width=0.32\textwidth]{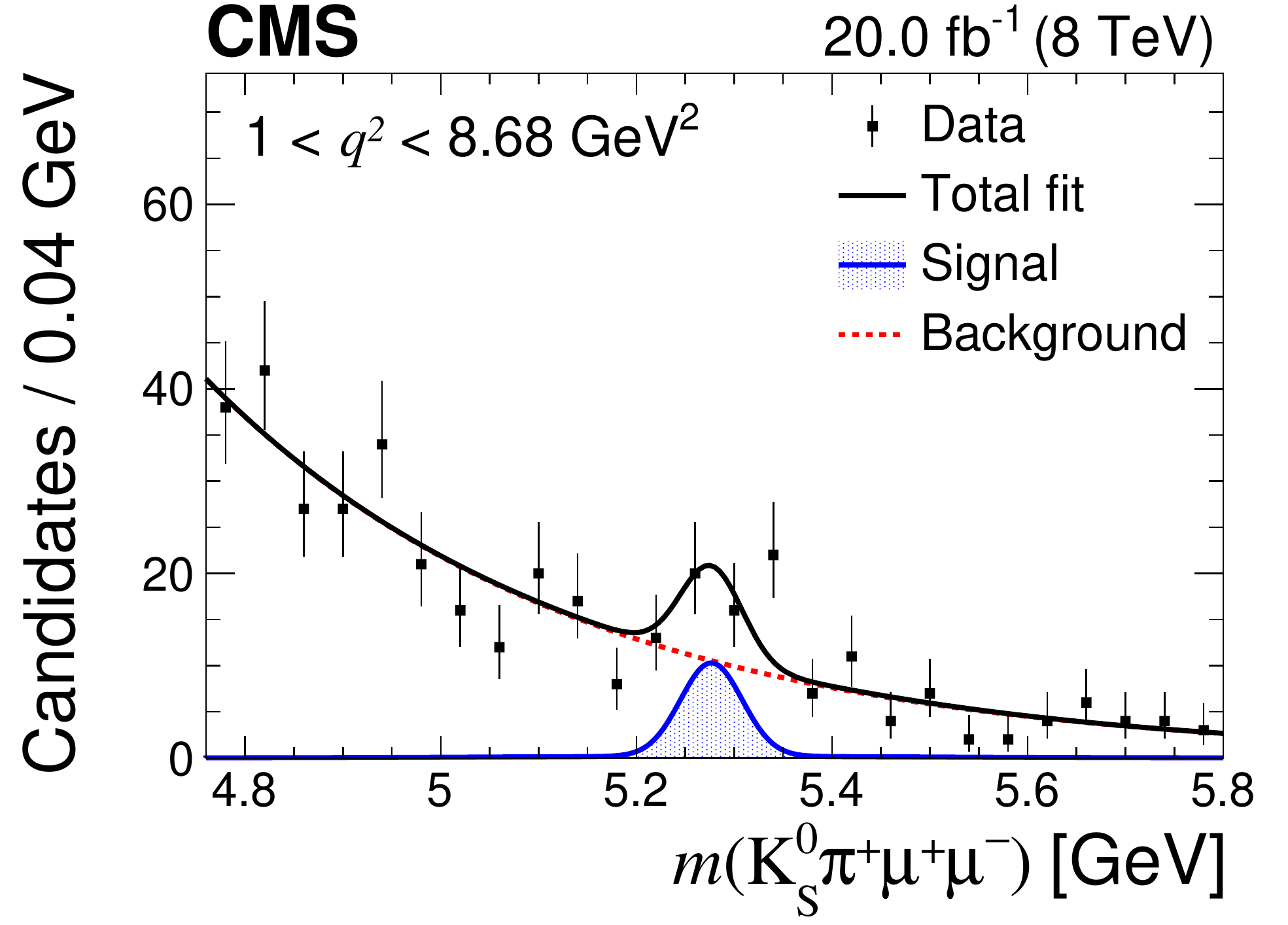}
    \includegraphics[width=0.32\textwidth]{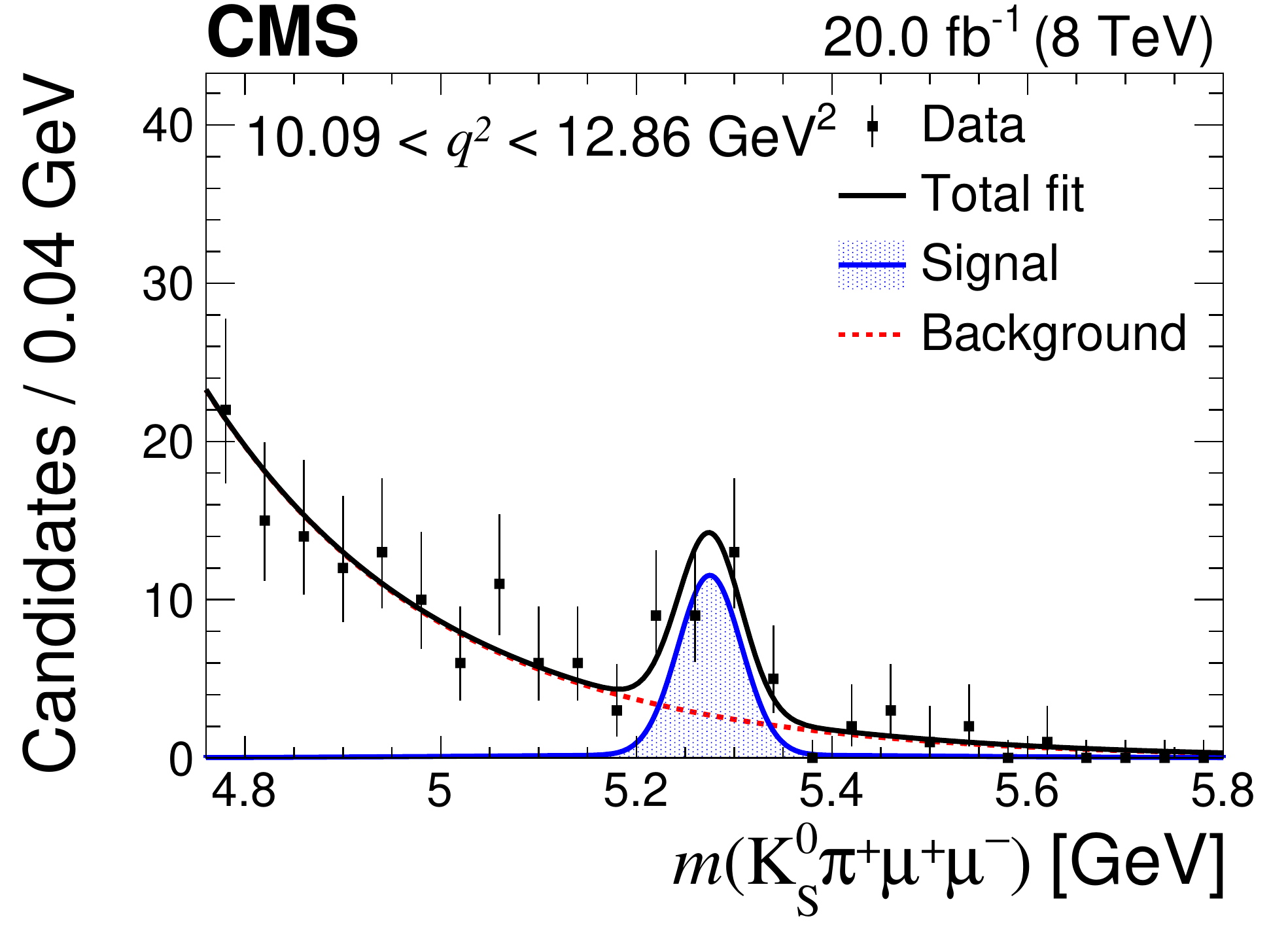}
    \includegraphics[width=0.32\textwidth]{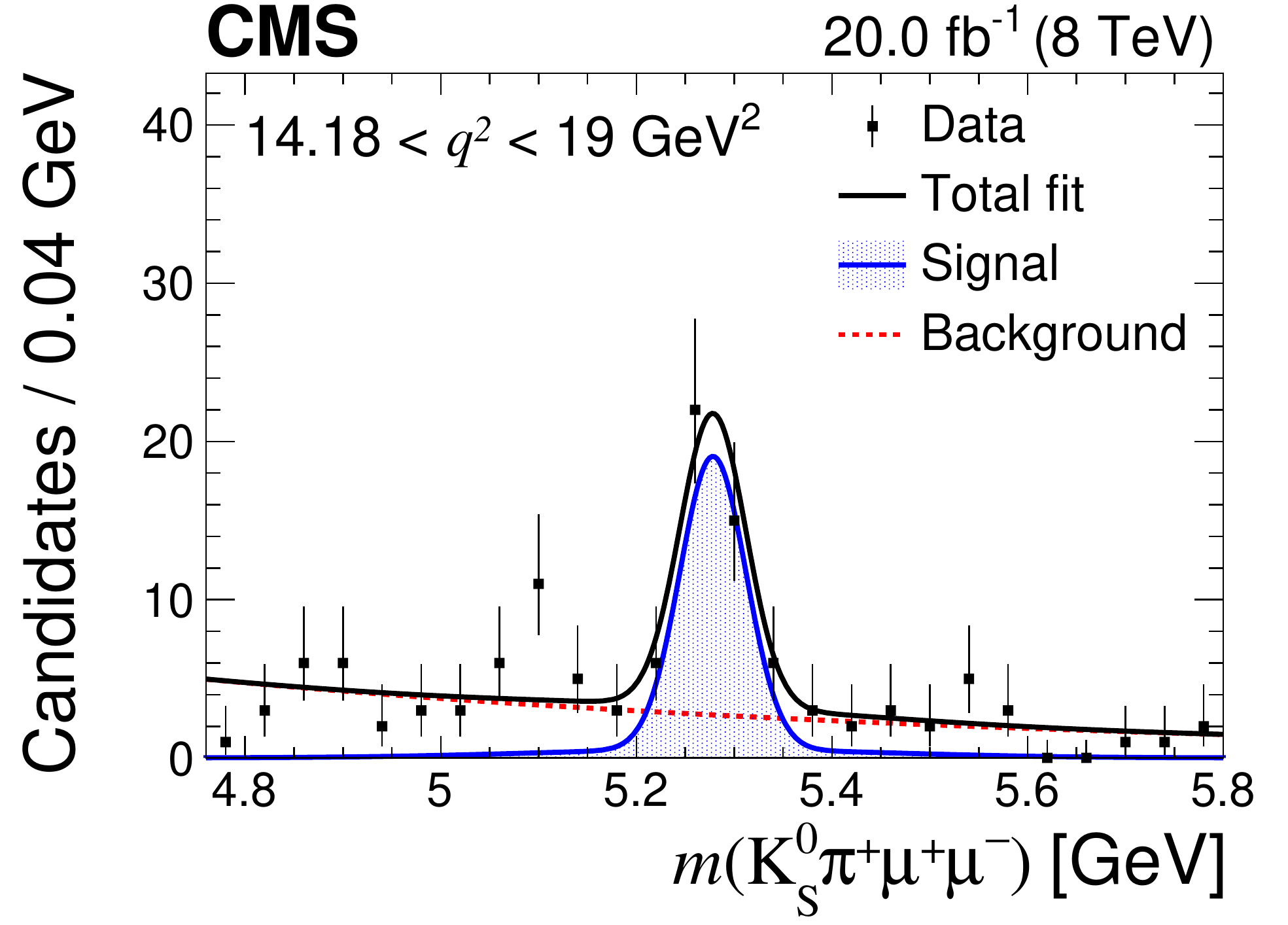} 
    \includegraphics[width=0.32\textwidth]{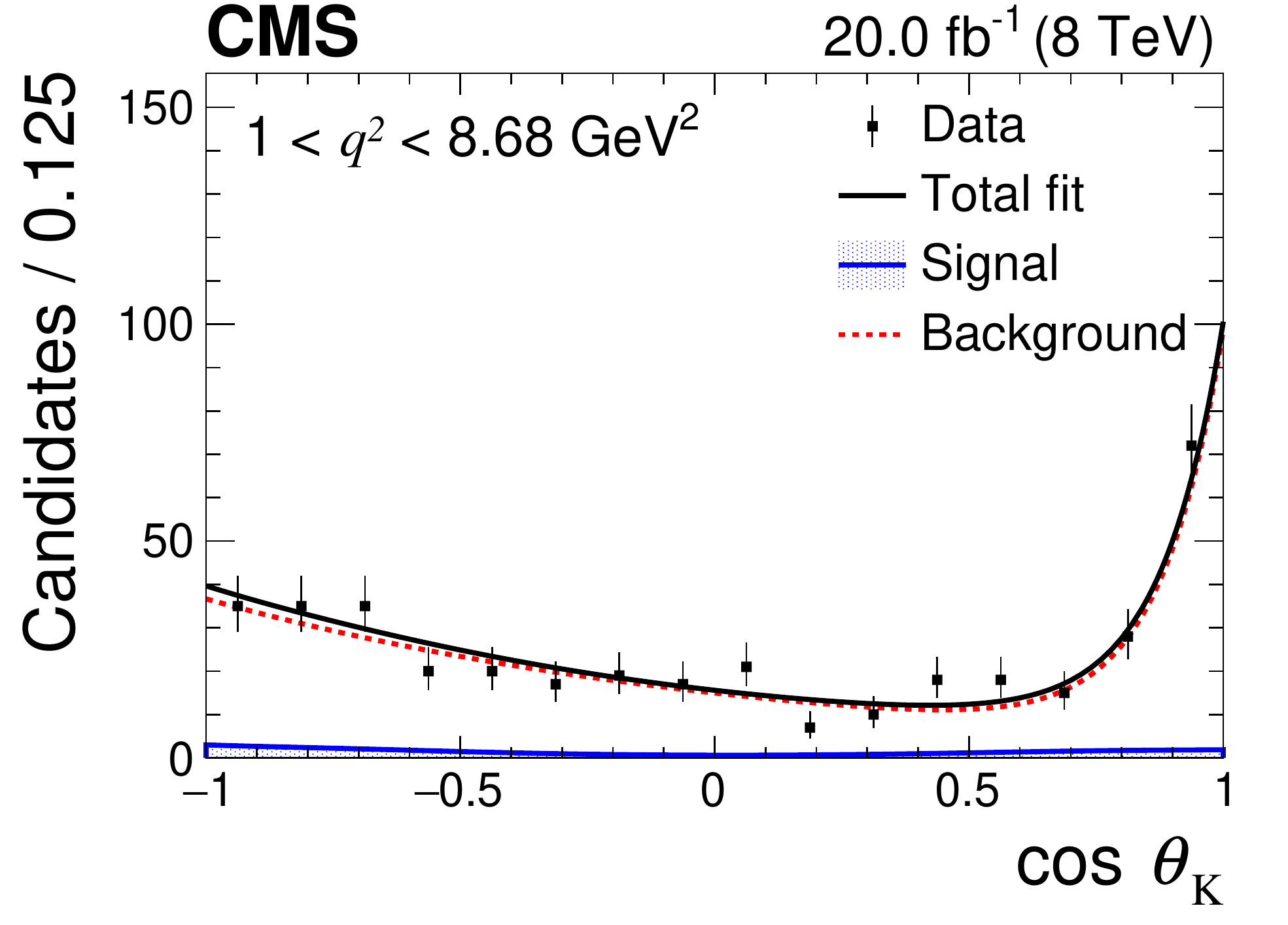}
    \includegraphics[width=0.32\textwidth]{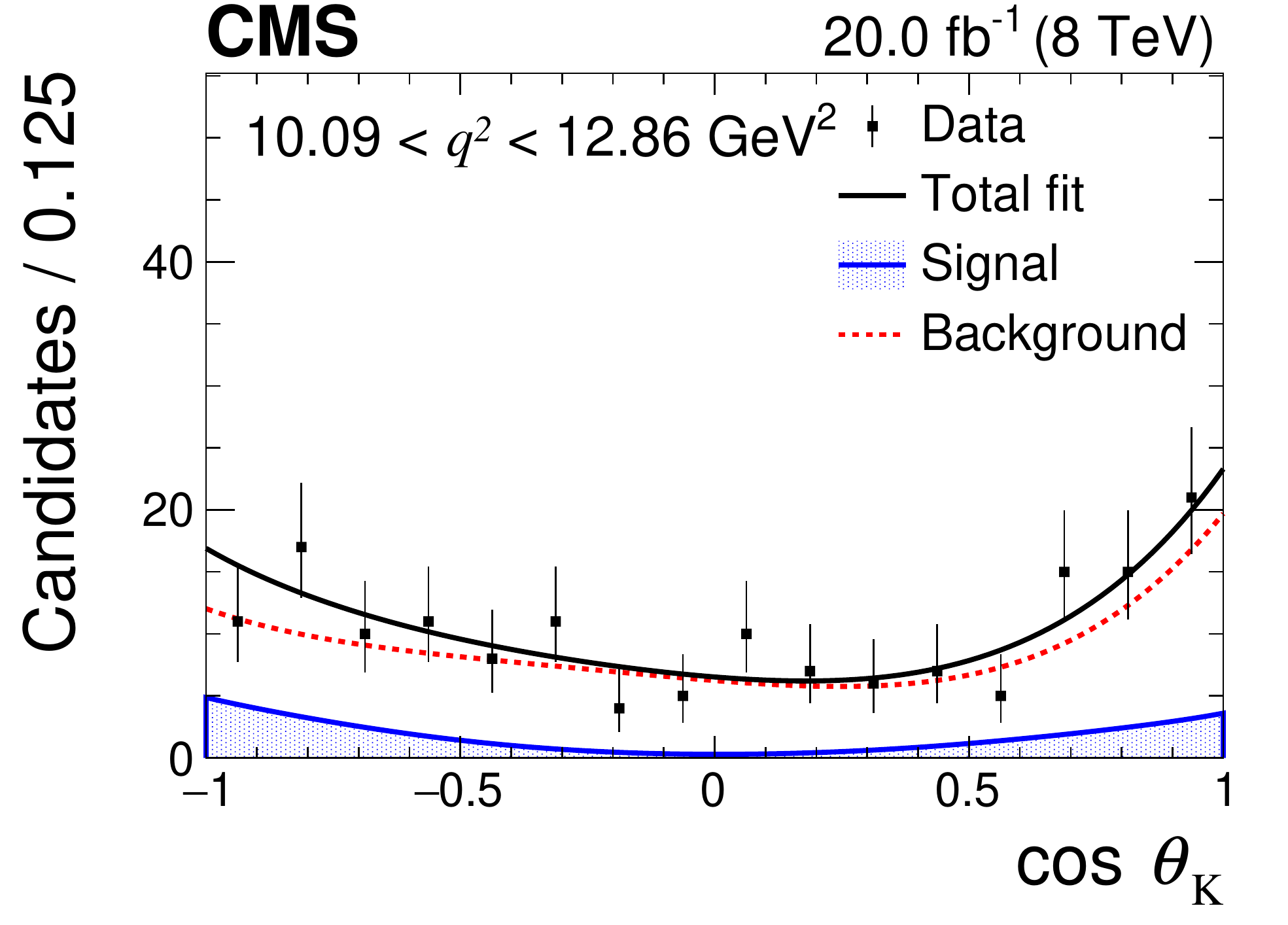}
    \includegraphics[width=0.32\textwidth]{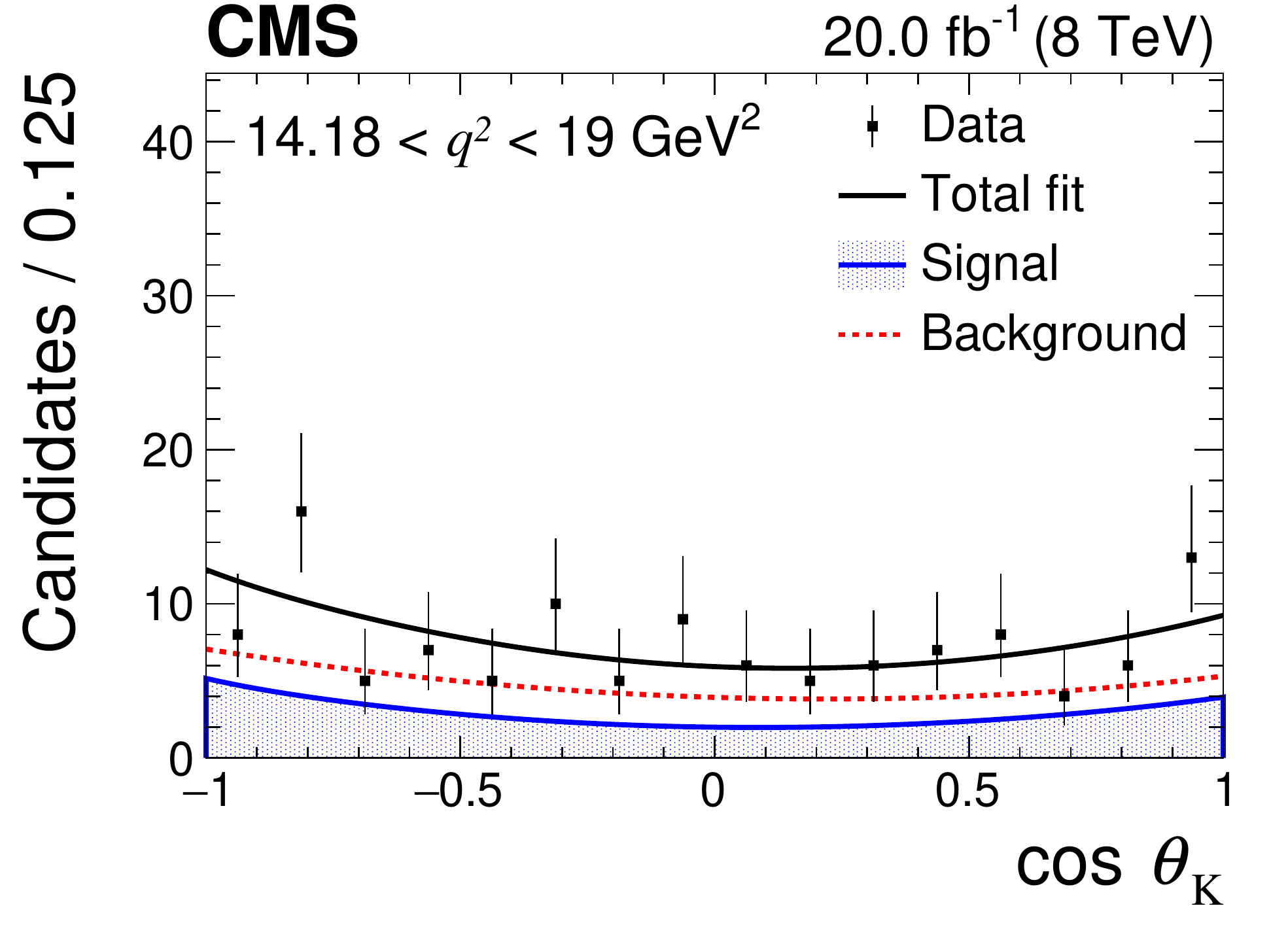} 
    \includegraphics[width=0.32\textwidth]{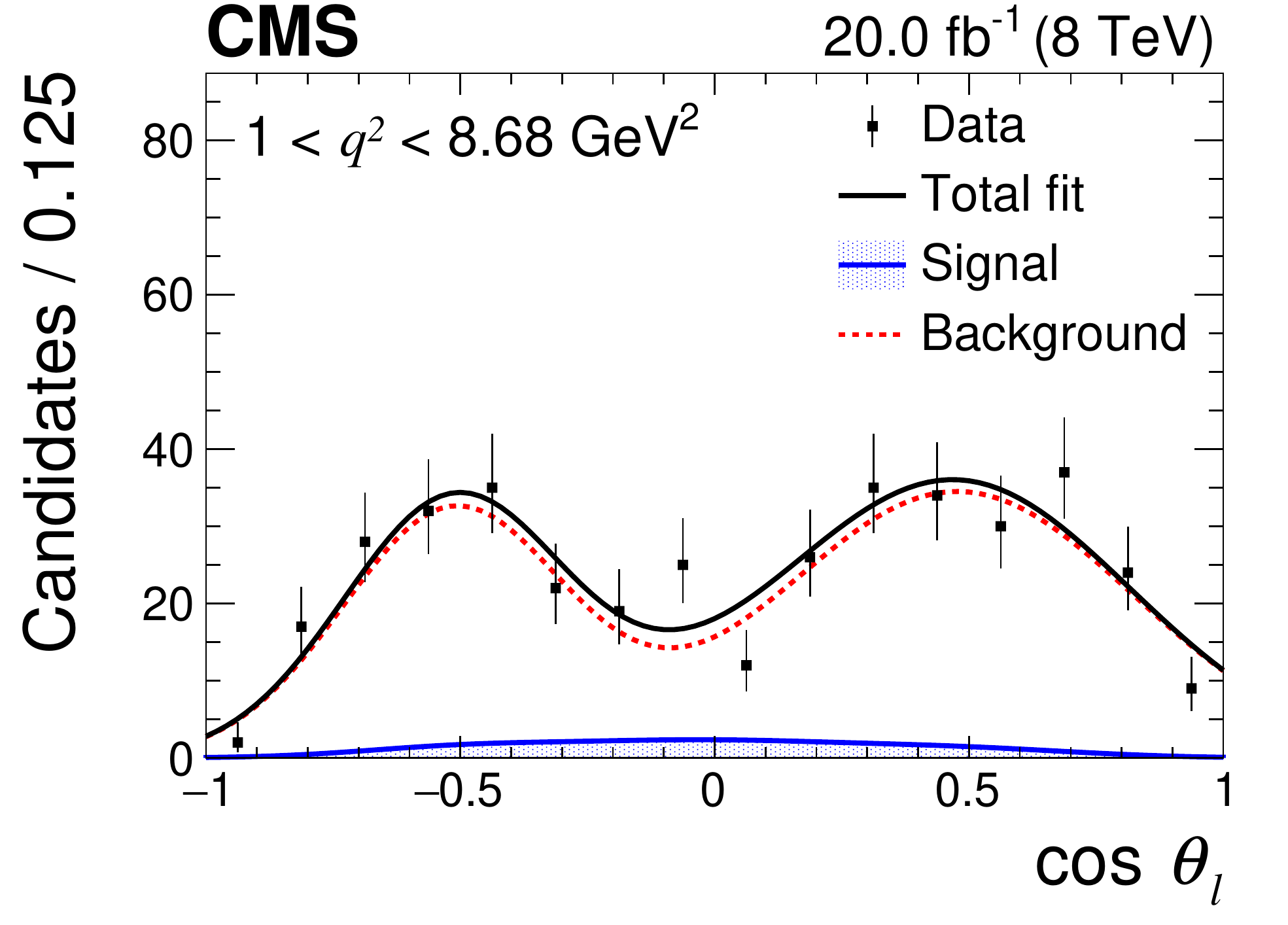}
    \includegraphics[width=0.32\textwidth]{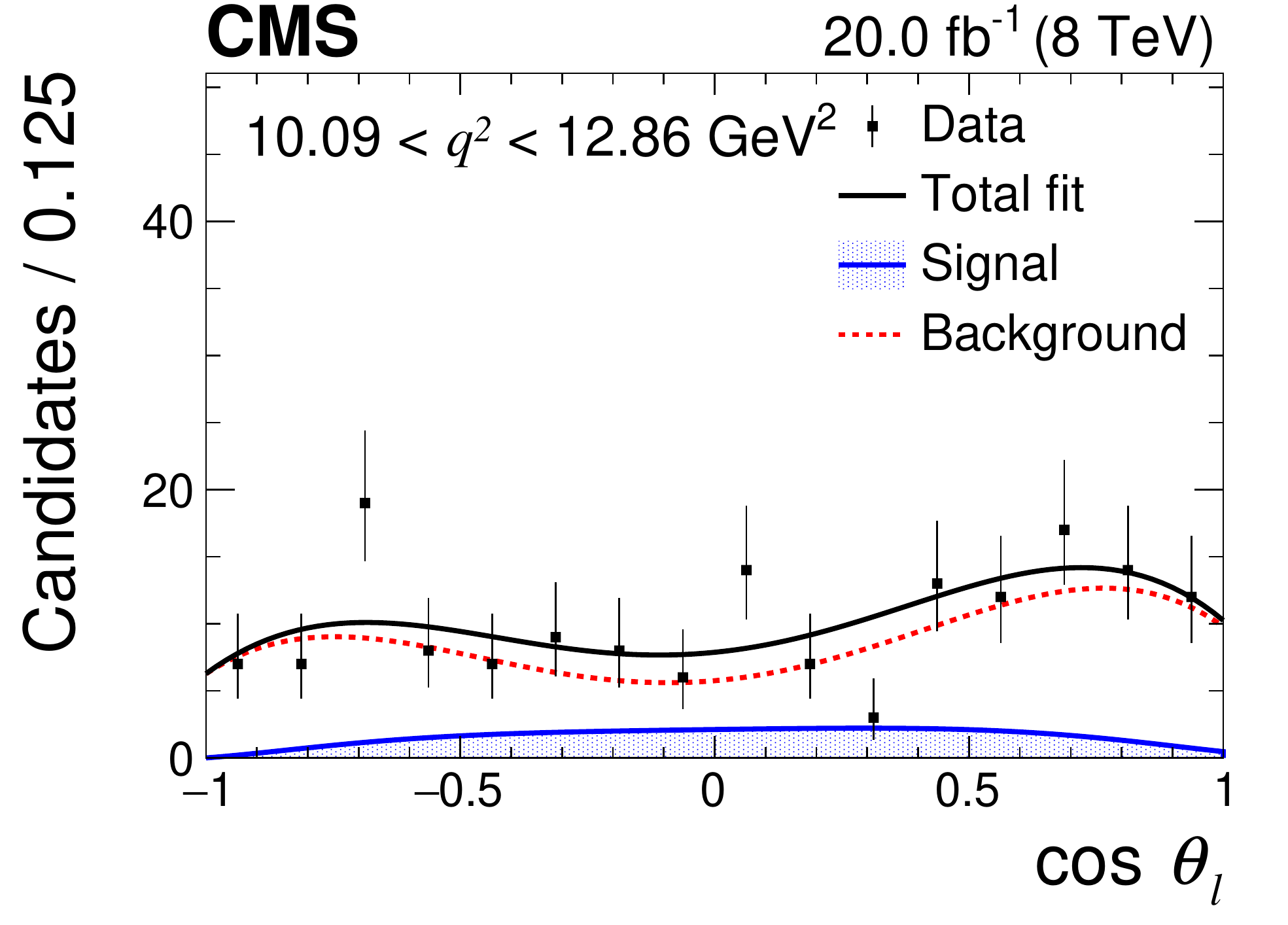}
    \includegraphics[width=0.32\textwidth]{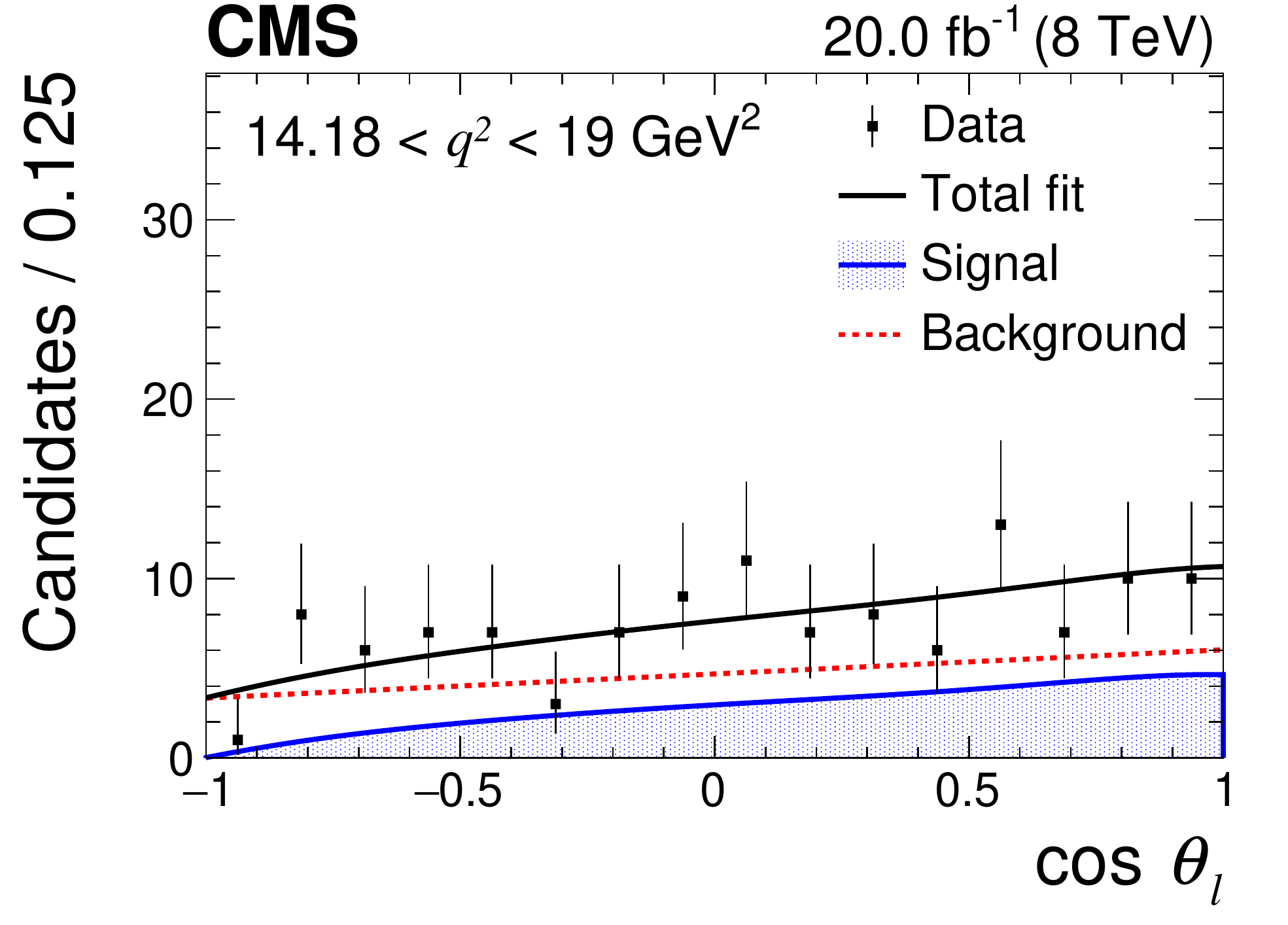}
    \caption{The $\PKzS\PGpp\PGmp\PGmm$ invariant mass (upper row),  $\cos\theta_\PK$ (middle row),  and $\cos\theta_\ell$ (lower row) 
    distributions for each \qq range is shown for data, along with the fit projections.
    The vertical bars on the data points indicate the statistical uncertainty. The filled areas, dashed lines, and
    solid lines represent the signal, background, and total contributions, respectively.}
       \label{fig:mass}
\end{figure}

\begin{figure}[htbp]
    \centering
    \includegraphics[width=0.32\textwidth]{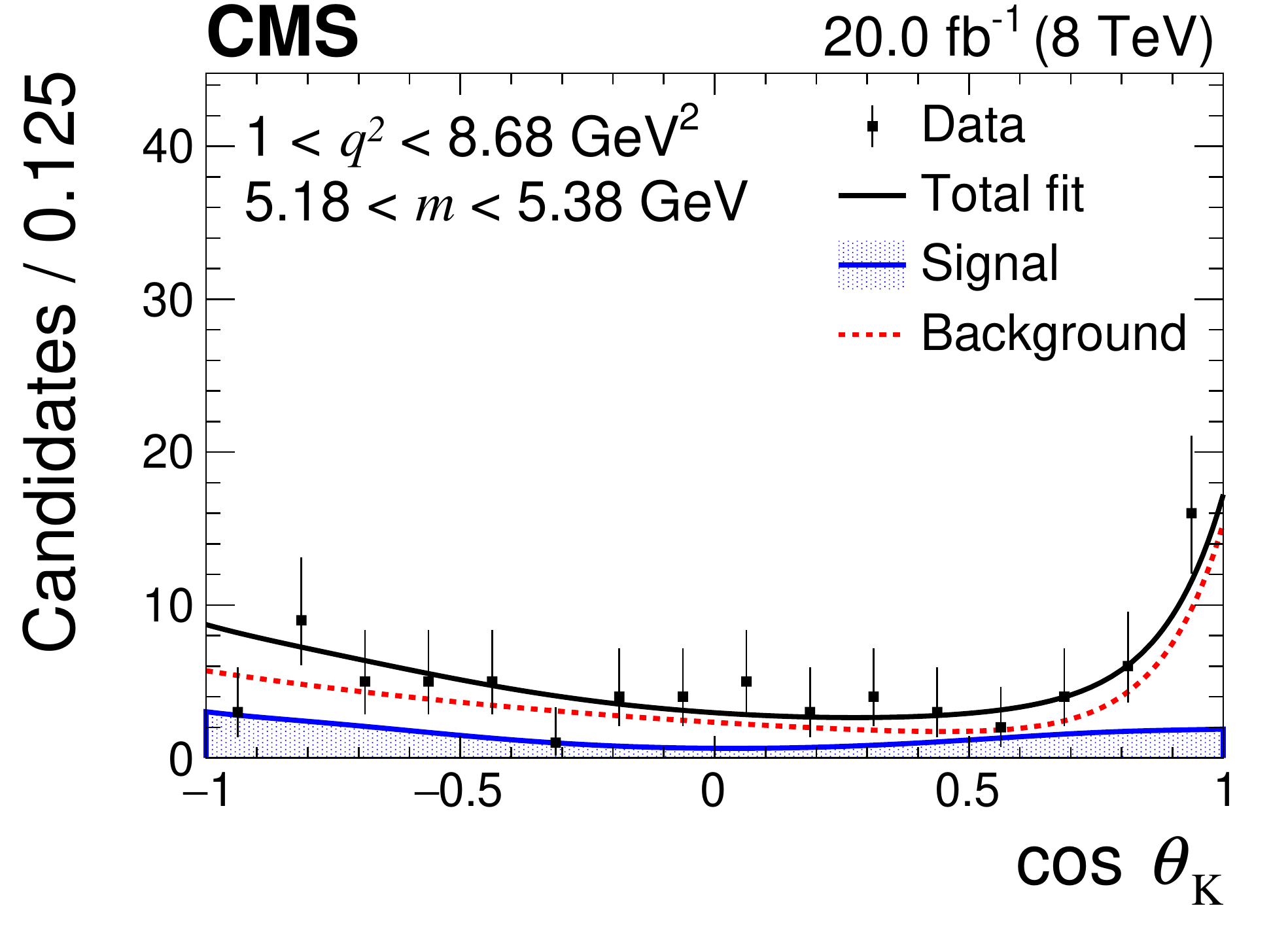}
    \includegraphics[width=0.32\textwidth]{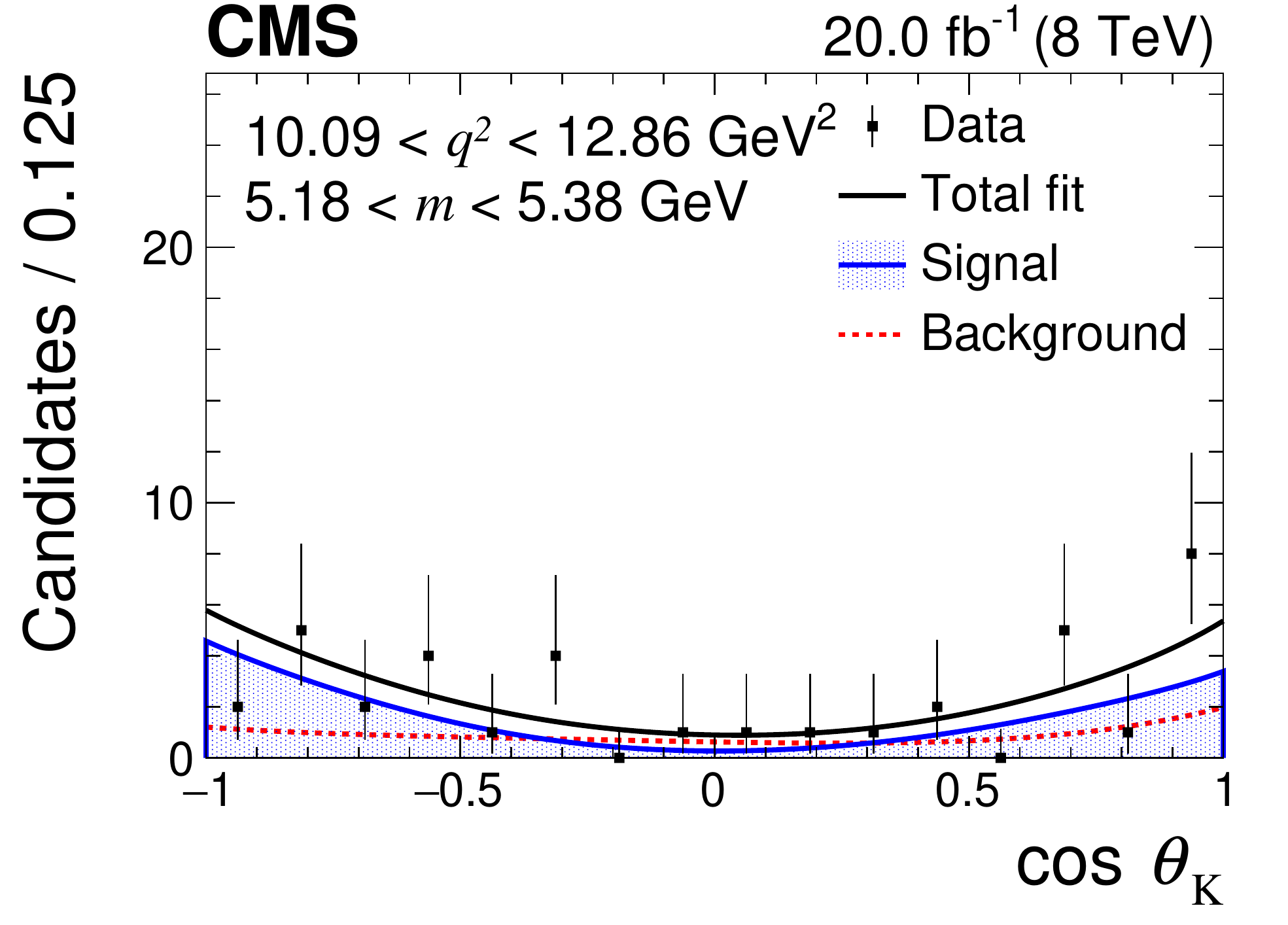}
    \includegraphics[width=0.32\textwidth]{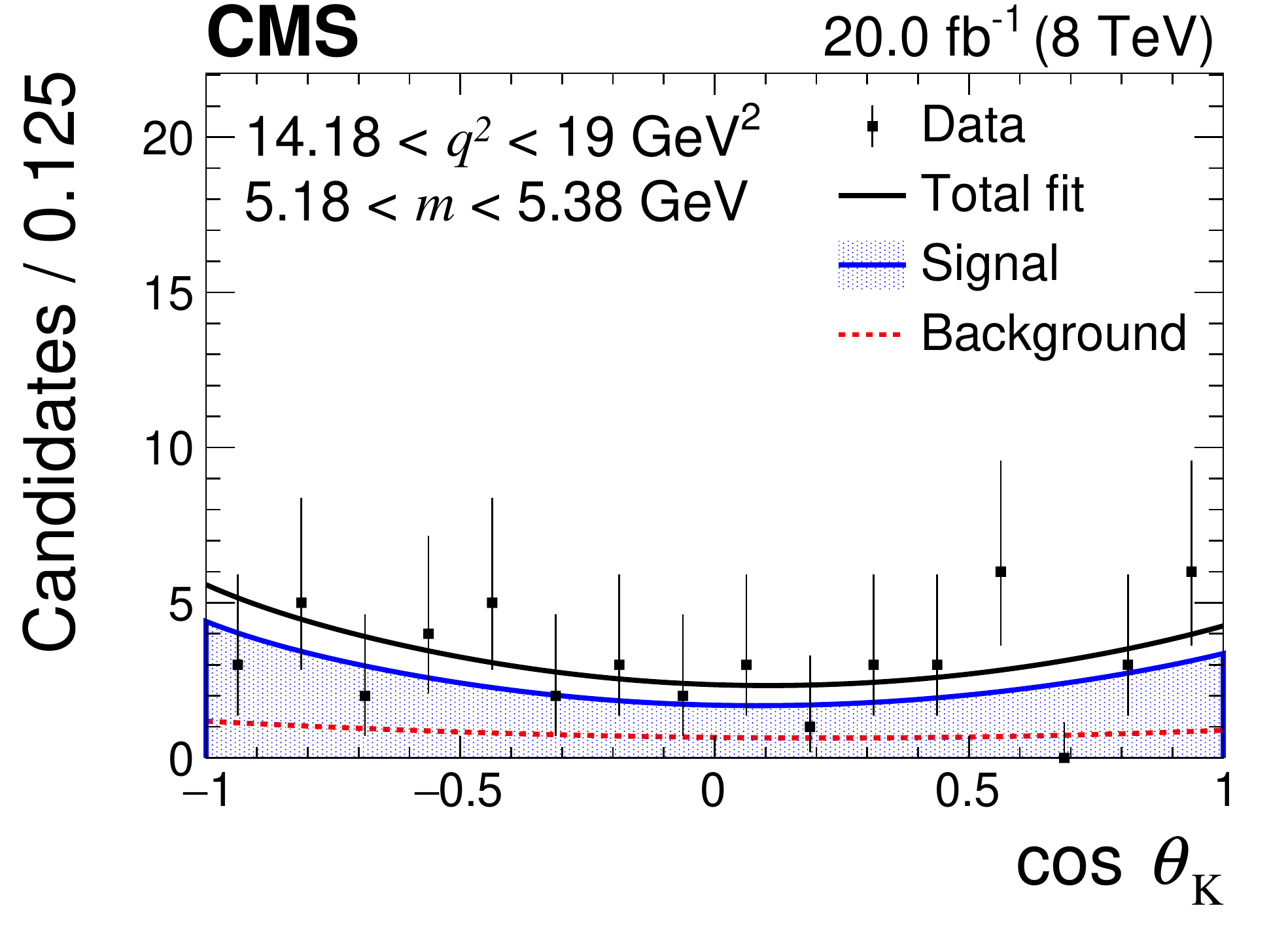}
    \includegraphics[width=0.32\textwidth]{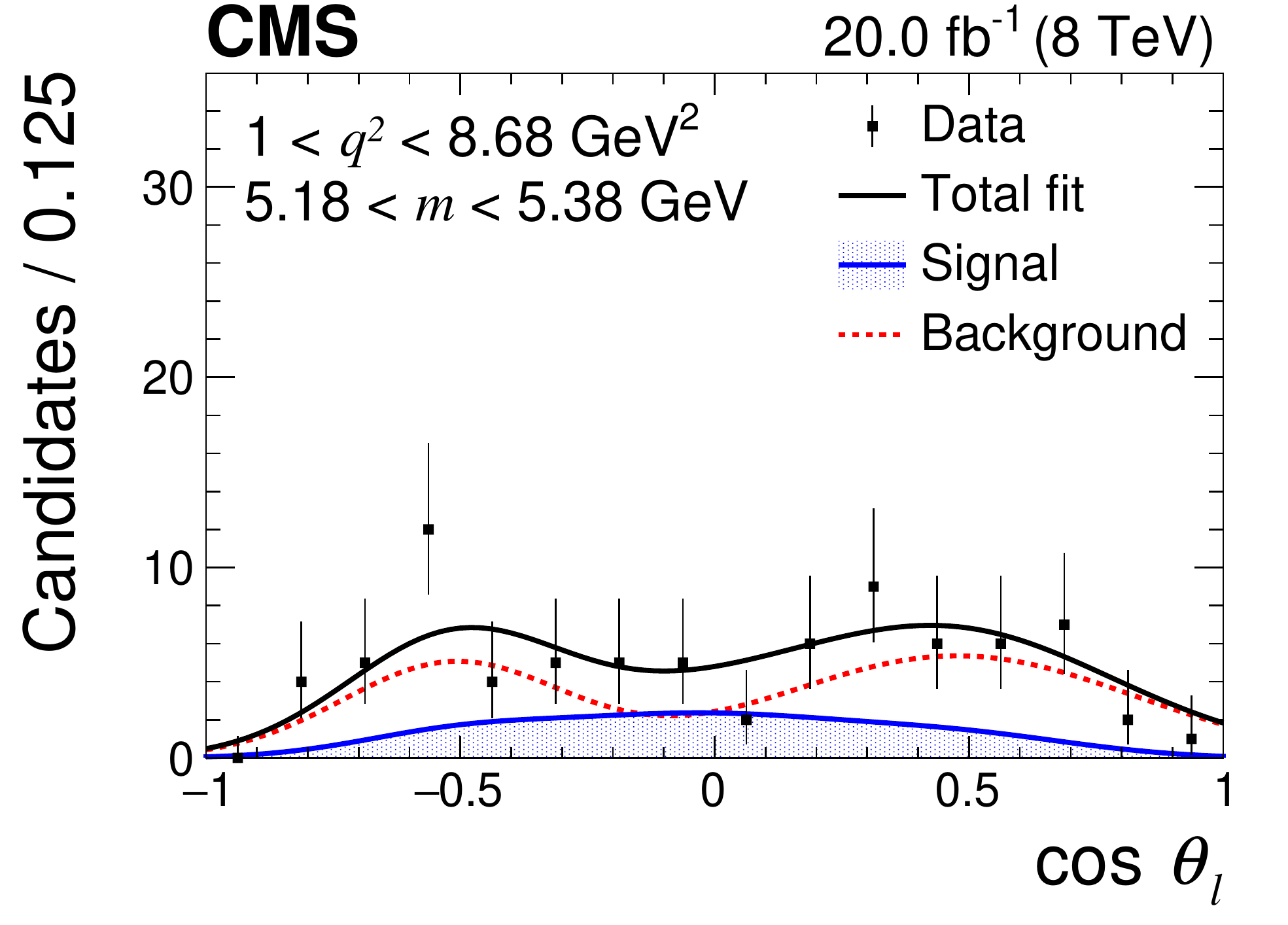}
    \includegraphics[width=0.32\textwidth]{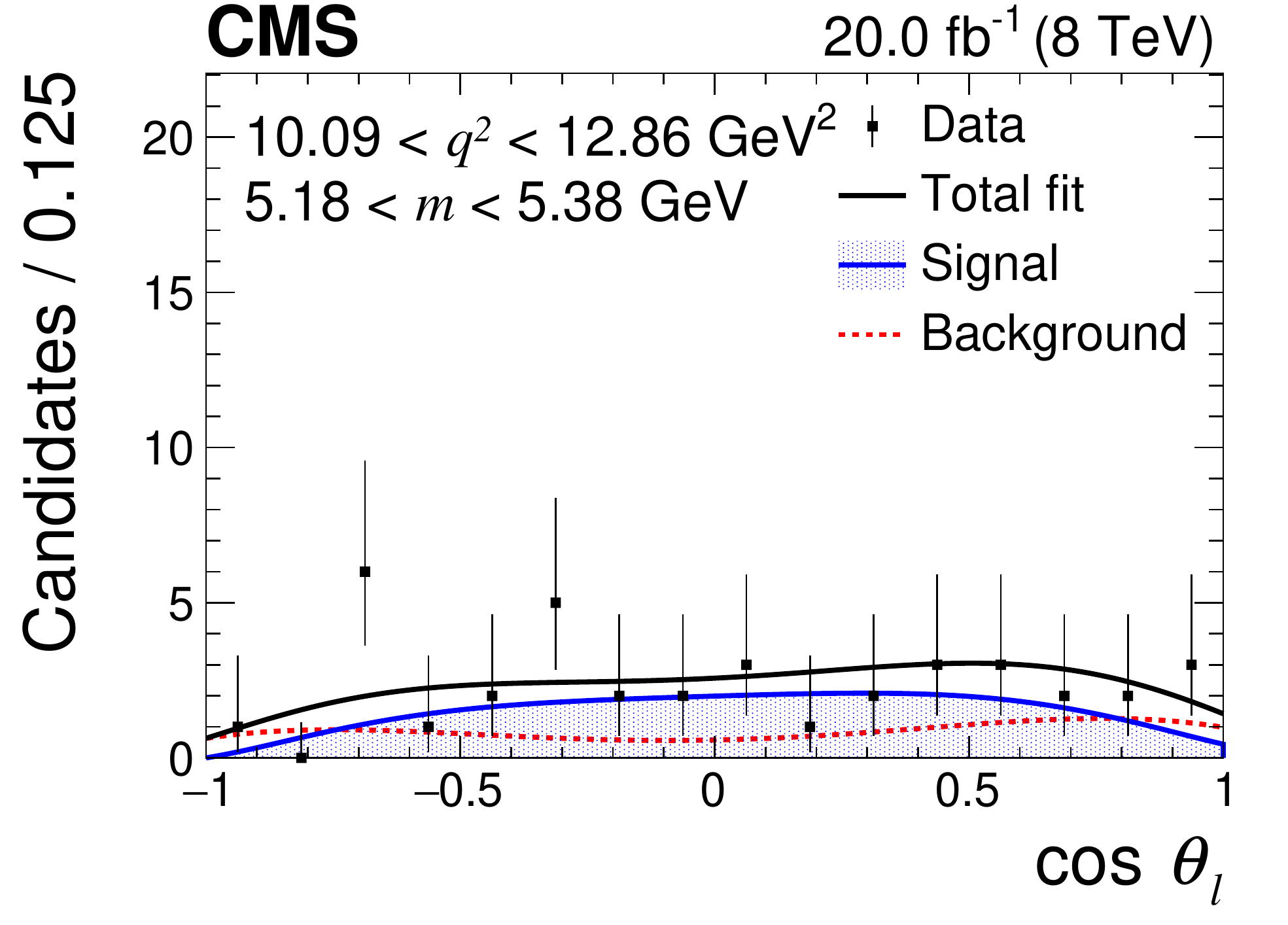}
    \includegraphics[width=0.32\textwidth]{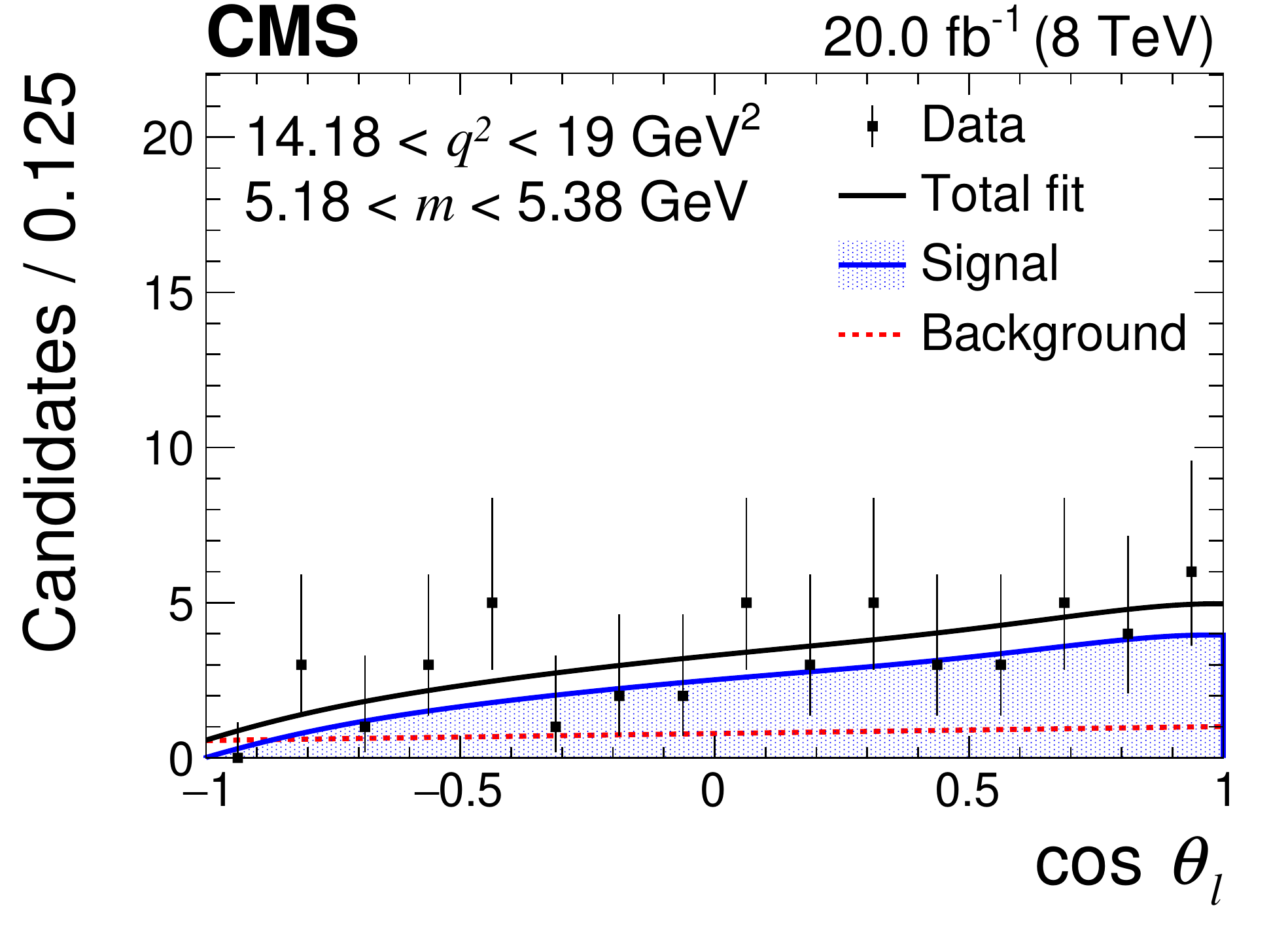}
    \caption{The $\cos\theta_\PK$ (upper row) and $\cos\theta_\ell$ (lower row) distributions for each \qq range is shown for data 
    in the invariant mass region $5.18<m<5.38\GeV$, along with the fit projections for the same region.
    The vertical bars on the data points indicate the statistical uncertainty. The filled areas, dashed lines, and
    solid lines represent the signal, background, and total contributions, respectively.}
       \label{fig:ang_SR}
\end{figure}

\begin{figure}[htbp]
    \centering
    \includegraphics[width=0.49\textwidth]{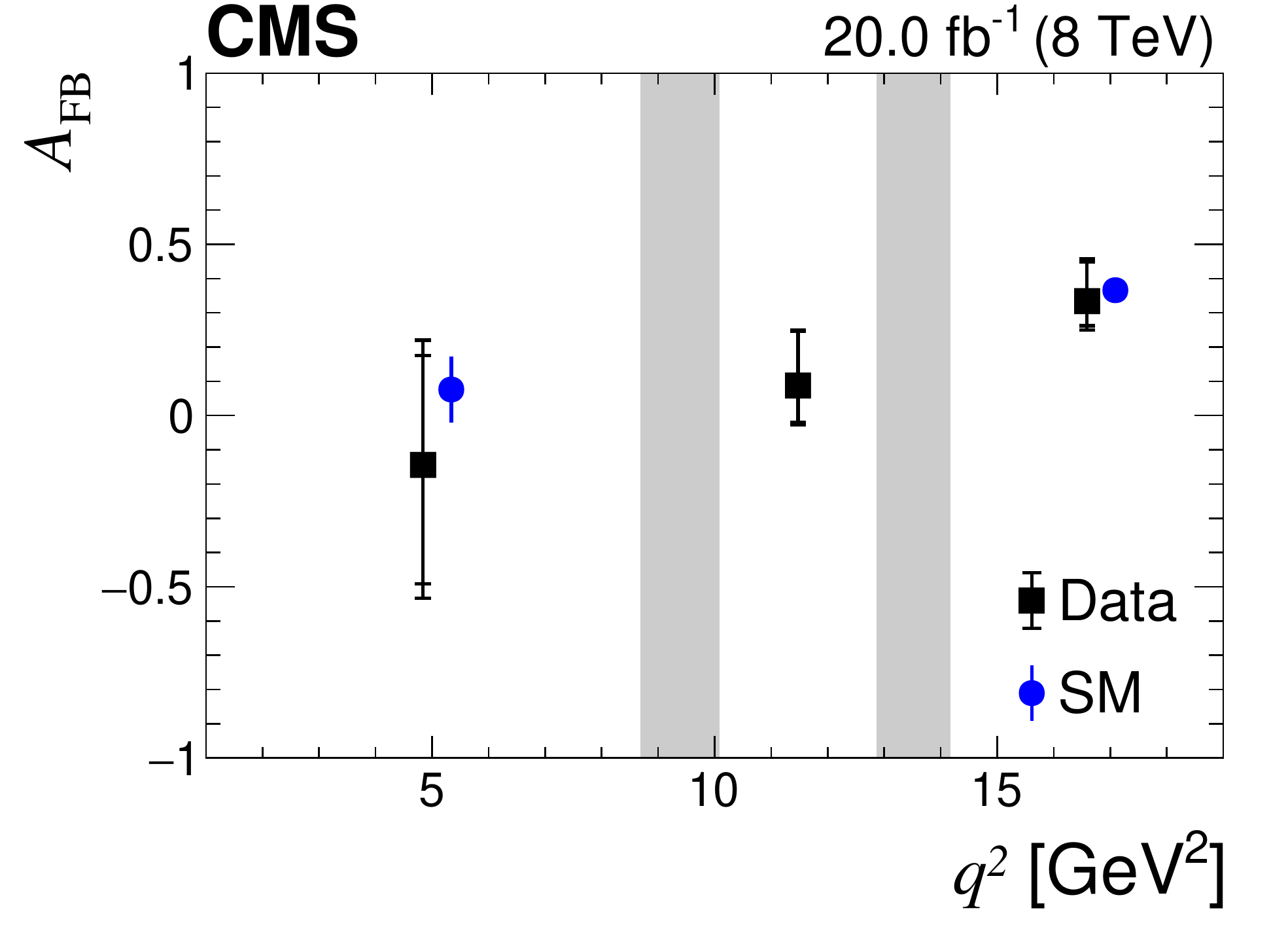}
    \includegraphics[width=0.49\textwidth]{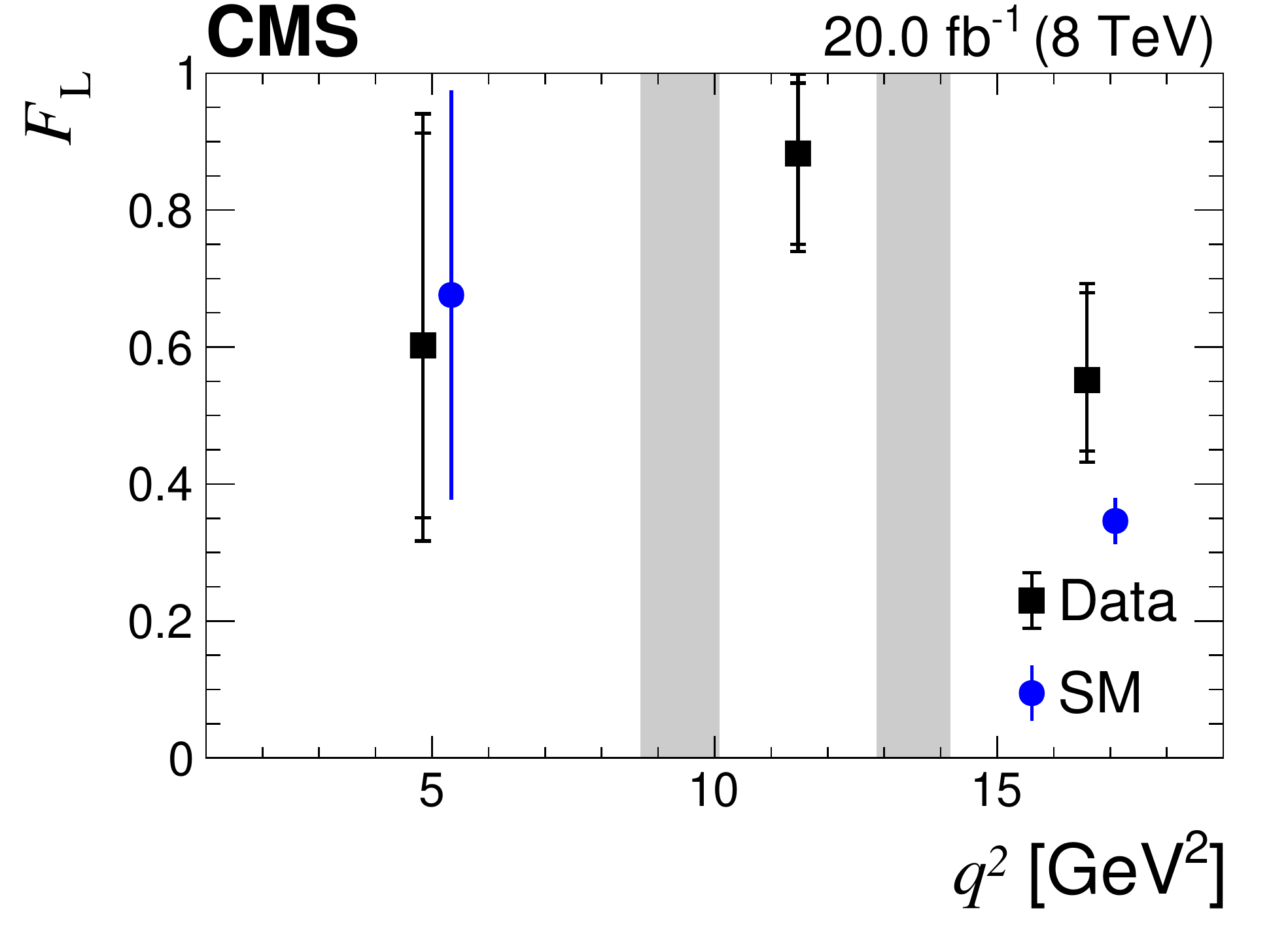}
    \caption{The measured values of \afb (left) and \fl (right) versus \qq for \Sig decays are shown with filled squares,
    centered on the \qq bin. The
    statistical (total) uncertainty is shown by inner (outer) vertical bars. The vertical shaded regions correspond to  
              the regions dominated by \JPS and \PSP decays. The SM predictions and associated uncertainties are shown by the filled
              circles and vertical bars, with the points slightly offset from the center of the \qq bin for clarity.}
    \label{fig:final}
\end{figure}

The results of the unbinned maximum likelihood fit are overlaid on the data in projections of $m$ (upper plots),
$\cos\theta_\PK$ (middle plots), and $\cos\theta_\ell$ (lower plots) for each \qq region in Fig.~\ref{fig:mass}.
The fitted values of $Y_S$, \afb, and \fl, along with their associated uncertainties, are given in Table~\ref{tab:results}
for each of the \qq bins.  In order to more clearly observe the signal features, the data and fit results are shown
versus the two angular variables in the invariant mass signal
region $5.18<m<5.38\GeV$ in Fig.~\ref{fig:ang_SR}. The fitted values of \afb and \fl are shown as a function
of \qq in Fig.~\ref{fig:final}, along with a SM prediction.
This prediction combines quantum chromodynamic factorization and soft collinear effective theory at large recoil 
with heavy-quark effective theory and lattice gauge theory at small recoil to separate hard physics (around the \PQb quark mass) 
from soft physics (around $\Lambda_\textrm{QCD}$)~\cite{Khodjamirian:2010vf, Horgan:2013hoa, Horgan:2015vla, Descotes-Genon:2014uoa}. While theoretical 
predictions are unavailable for the region between the \JPsi and \Pgy meson masses ($10.09 < \qq < 12.86\GeV^{2}$), the SM prediction agrees
with the experimental results for the other \qq bins, indicating no evidence of contributions from physics beyond the SM.

\begin{table}[htbp]
  \centering
  \topcaption{The $Y_S$, \afb, and \fl values from the fit for each \qq range. The first uncertainty is statistical and the second is systematic.
  \label{tab:results}} 
  \renewcommand{\arraystretch}{1.3}
    \begin{tabular}{r@{~}c@{~}lccc}
      \multicolumn{3}{c}{\qq (${\GeVns}^2$)} & $Y_S$ & \afb & \fl \\
        \hline
        1 & -- & 8.68 & $22.1 \pm 8.1$ & $-0.14^{\,+0.32}_{\,{-}0.35} \pm 0.17$ & $0.60^{\,+0.31}_{\,{-}0.25} \pm 0.13$ \\
        10.09&  -- &12.86 & $25.9 \pm 6.3$ & $\;\;\;0.09^{\,+0.16}_{\,{-}0.11} \pm 0.04$ & $0.88^{\,+0.10}_{\,{-}0.13} \pm 0.05$ \\
        14.18& -- & 19 & $45.1 \pm 8.0$ & $\;\;\;0.33^{\,+0.11}_{\,{-}0.07} \pm 0.05$ & $0.55^{\,+0.13}_{\,{-}0.10} \pm 0.06$ \\
    \end{tabular}
\end{table}

\section{Summary}
The first angular analysis of the exclusive decay $\PBp\to \Kstmass \Pgmp \Pgmm$, including the charge-conjugate state, has been 
performed using a sample of proton-proton collisions at a center-of-mass energy of 8\TeV.  The data were collected 
with the CMS detector in 2012 at the LHC, and correspond to an integrated luminosity of $20.0\fbinv$.  For each bin 
of the dimuon invariant mass squared (\qq), a three-dimensional unbinned maximum likelihood fit is performed on the distributions
of the $\PKstP{892}^{+}\Pgmp\Pgmm$ invariant mass and two decay angles. The muon forward-backward asymmetry, \afb,
and the $\Kstmass$ longitudinal polarization fraction, \fl, are extracted from the fit in bins of \qq 
and found to be consistent with a standard model prediction.

\begin{acknowledgments}
We thank J.\ Matias for providing the theoretical values of \afb and \fl used for comparisons with our measurements. 

We congratulate our colleagues in the CERN accelerator departments for the excellent performance of the LHC and thank the technical and administrative staffs at CERN and at other CMS institutes for their contributions to the success of the CMS effort. In addition, we gratefully acknowledge the computing centers and personnel of the Worldwide LHC Computing Grid for delivering so effectively the computing infrastructure essential to our analyses. Finally, we acknowledge the enduring support for the construction and operation of the LHC and the CMS detector provided by the following funding agencies: BMBWF and FWF (Austria); FNRS and FWO (Belgium); CNPq, CAPES, FAPERJ, FAPERGS, and FAPESP (Brazil); MES (Bulgaria); CERN; CAS, MoST, and NSFC (China); COLCIENCIAS (Colombia); MSES and CSF (Croatia); RIF (Cyprus); SENESCYT (Ecuador); MoER, ERC PUT and ERDF (Estonia); Academy of Finland, MEC, and HIP (Finland); CEA and CNRS/IN2P3 (France); BMBF, DFG, and HGF (Germany); GSRT (Greece); NKFIA (Hungary); DAE and DST (India); IPM (Iran); SFI (Ireland); INFN (Italy); MSIP and NRF (Republic of Korea); MES (Latvia); LAS (Lithuania); MOE and UM (Malaysia); BUAP, CINVESTAV, CONACYT, LNS, SEP, and UASLP-FAI (Mexico); MOS (Montenegro); MBIE (New Zealand); PAEC (Pakistan); MSHE and NSC (Poland); FCT (Portugal); JINR (Dubna); MON, RosAtom, RAS, RFBR, and NRC KI (Russia); MESTD (Serbia); SEIDI, CPAN, PCTI, and FEDER (Spain); MOSTR (Sri Lanka); Swiss Funding Agencies (Switzerland); MST (Taipei); ThEPCenter, IPST, STAR, and NSTDA (Thailand); TUBITAK and TAEK (Turkey); NASU (Ukraine); STFC (United Kingdom); DOE and NSF (USA).

\hyphenation{Rachada-pisek} Individuals have received support from the Marie-Curie program and the European Research Council and Horizon 2020 Grant, contract Nos.\ 675440, 724704, 752730, and 765710 (European Union); the Leventis Foundation; the A.P.\ Sloan Foundation; the Alexander von Humboldt Foundation; the Belgian Federal Science Policy Office; the Fonds pour la Formation \`a la Recherche dans l'Industrie et dans l'Agriculture (FRIA-Belgium); the Agentschap voor Innovatie door Wetenschap en Technologie (IWT-Belgium); the F.R.S.-FNRS and FWO (Belgium) under the ``Excellence of Science -- EOS" -- be.h project n.\ 30820817; the Beijing Municipal Science \& Technology Commission, No. Z191100007219010; the Ministry of Education, Youth and Sports (MEYS) of the Czech Republic; the Deutsche Forschungsgemeinschaft (DFG) under Germany's Excellence Strategy -- EXC 2121 ``Quantum Universe" -- 390833306; the Lend\"ulet (``Momentum") Program and the J\'anos Bolyai Research Scholarship of the Hungarian Academy of Sciences, the New National Excellence Program \'UNKP, the NKFIA research grants 123842, 123959, 124845, 124850, 125105, 128713, 128786, and 129058 (Hungary); the Council of Science and Industrial Research, India; the HOMING PLUS program of the Foundation for Polish Science, cofinanced from European Union, Regional Development Fund, the Mobility Plus program of the Ministry of Science and Higher Education, the National Science Center (Poland), contracts Harmonia 2014/14/M/ST2/00428, Opus 2014/13/B/ST2/02543, 2014/15/B/ST2/03998, and 2015/19/B/ST2/02861, Sonata-bis 2012/07/E/ST2/01406; the National Priorities Research Program by Qatar National Research Fund; the Ministry of Science and Higher Education, project no. 02.a03.21.0005 (Russia); the Tomsk Polytechnic University Competitiveness Enhancement Program; the Programa Estatal de Fomento de la Investigaci{\'o}n Cient{\'i}fica y T{\'e}cnica de Excelencia Mar\'{\i}a de Maeztu, grant MDM-2015-0509 and the Programa Severo Ochoa del Principado de Asturias; the Thalis and Aristeia programs cofinanced by EU-ESF and the Greek NSRF; the Rachadapisek Sompot Fund for Postdoctoral Fellowship, Chulalongkorn University and the Chulalongkorn Academic into Its 2nd Century Project Advancement Project (Thailand); the Kavli Foundation; the Nvidia Corporation; the SuperMicro Corporation; the Welch Foundation, contract C-1845; and the Weston Havens Foundation (USA).
\end{acknowledgments}

\bibliography{auto_generated}  
\cleardoublepage \appendix\section{The CMS Collaboration \label{app:collab}}\begin{sloppypar}\hyphenpenalty=5000\widowpenalty=500\clubpenalty=5000\vskip\cmsinstskip
\textbf{Yerevan Physics Institute, Yerevan, Armenia}\\*[0pt]
A.M.~Sirunyan$^{\textrm{\dag}}$, A.~Tumasyan
\vskip\cmsinstskip
\textbf{Institut f\"{u}r Hochenergiephysik, Wien, Austria}\\*[0pt]
W.~Adam, T.~Bergauer, M.~Dragicevic, A.~Escalante~Del~Valle, R.~Fr\"{u}hwirth\cmsAuthorMark{1}, M.~Jeitler\cmsAuthorMark{1}, N.~Krammer, L.~Lechner, D.~Liko, I.~Mikulec, F.M.~Pitters, N.~Rad, J.~Schieck\cmsAuthorMark{1}, R.~Sch\"{o}fbeck, M.~Spanring, S.~Templ, W.~Waltenberger, C.-E.~Wulz\cmsAuthorMark{1}, M.~Zarucki
% ...
\vskip\cmsinstskip
\dag: Deceased\\
1:  Also at Vienna University of Technology, Vienna, Austria\\
% ...
\end{sloppypar}
\end{document}